\documentclass[journal, comsoc]{IEEEtran}
\usepackage{cite}
\usepackage{amsmath,amssymb,amsfonts}
\usepackage{algorithmic}
\usepackage{graphicx,color}
\usepackage{textcomp}

\usepackage{bm,amsthm}
\usepackage{tikz}
\usepackage{pgfplots}
\pgfplotsset{compat=newest}
\usetikzlibrary{calc, shapes.geometric, shapes.misc, arrows, matrix, fit, decorations.markings}
\usepackage[absolute, overlay]{textpos}
\usepackage{scalerel}
\usepackage{hyperref}

\newcommand{\reffig}[1]{Fig.~\ref{#1}}                  %
\newcommand{\reftable}[1]{Tab.~\ref{#1}}                %
\newcommand{\refsec}[1]{Sec.~\ref{#1}}                  %
\newcommand{\aprob}{\ensuremath{q}} 				    %
\newcommand{\param}{\ensuremath{\theta}} 			    %
\newcommand{\params}{\ensuremath{\bog{\param}}} 	    %
\newcommand{\viparams}{\ensuremath{\bog{\varphi}}} 		%
\newcommand{\tset}{\ensuremath{\mathcal{D}}} 		    %
\newcommand{\Nsamp}{\ensuremath{N}} 				    %
\newcommand{\modelfeatures}{\ensuremath{\phi}}          %
\newcommand{\idxn}{\ensuremath{n}}                      %
\newcommand{\idxi}{\ensuremath{i}}                      %

\newcommand{\bo}[1]{\ensuremath{\mathbf{#1}}}           %
\newcommand{\bog}[1]{\ensuremath{\bm{#1}}}              %
\newcommand{\txt}[1]{\ensuremath{\text{#1}}}            %
\newcommand{\mathtxt}[1]{\ensuremath{\mathrm{#1}}}      %
\newcommand{\eqpoint}{\ensuremath{\,.}}                 %
\newcommand{\eqcomma}{\ensuremath{\,,}}                 %

\DeclareMathOperator*{\armax}{arg\,max}                                     %
\DeclareMathOperator*{\armin}{arg\,min}                                     %
\newcommand{\argmax}[1][\cdot]{\ensuremath{\underset{#1}{\armax}\ }}        %
\newcommand{\argmin}[1][\cdot]{\ensuremath{\underset{#1}{\armin}\ }}        %

\newcommand{\lpa}{\ensuremath{\left(}}
\newcommand{\rpa}{\ensuremath{\right)}}
\newcommand{\lbb}{\ensuremath{\left[}}
\newcommand{\rbb}{\ensuremath{\right]}}
\newcommand{\lcb}{\ensuremath{\left\{}}
\newcommand{\rcb}{\ensuremath{\right\}}}

\newcommand{\summ}[2][]{\ensuremath{\sum \limits_{#2}^{#1}}}                %

\DeclareMathOperator{\E}{E} %
\newcommand{\eval}[2][\cdot]{\ensuremath{\E_{#1}\negthinspace\lbb#2\rbb}} 	%
\newcommand{\evalnb}[2][\cdot]{\ensuremath{\E_{#1}\negthinspace[#2]}} 		%
\newcommand{\dkl}[2]{\ensuremath{D_{\mathtxt{KL}}\lpa#1 \parallel #2\rpa}} 	%
\newcommand{\he}{\ensuremath{\mathcal{H}}}									%
\newcommand{\HE}[1][\cdot]{\ensuremath{\he\lpa#1\rpa}} 						%
\newcommand{\HEN}[2][]{\ensuremath{\he_{#1}\lpa#2\rpa}} 					%

\DeclareMathOperator{\dg}{diag}                                         %
\newcommand{\diag}[1]{\ensuremath{\dg\lcb #1\rcb}}                      %

\newcommand{\rv}{RV}                                                    %
\newcommand{\Nb}{\ensuremath{N_{\mathtxt{b}}}}                      %
\newcommand{\Ne}{\ensuremath{N_{\mathtxt{e}}}}                      %
\newcommand{\NL}{\ensuremath{N_{\mathtxt{L}}}}              %
\newcommand{\Nclass}{\ensuremath{M}}                        %
\newcommand{\xvar}{\ensuremath{x}}                          %
\newcommand{\xvec}{\ensuremath{\bo{x}}}                     %
\newcommand{\yvar}{\ensuremath{y}}                          %
\newcommand{\yvec}{\ensuremath{\bo{\yvar}}}                 %
\newcommand{\prob}{\ensuremath{p}}                          %
\DeclareMathOperator*{\ohfun}{one-hot}                      %
\newcommand{\ohfunc}{\ensuremath{\ohfun}}                   %
\newcommand{\indc}{\ensuremath{k}}                          %
\newcommand{\indi}{\ensuremath{i}}                          %
\newcommand{\plvar}{\ensuremath{a}}                         %
\newcommand{\plvec}{\ensuremath{\bo{\plvar}}}               %
\newcommand{\const}{\ensuremath{c}}                             %

\newcommand{\hidlayel}{\ensuremath{v}}							%
\newcommand{\hidlay}{\ensuremath{\bo{\hidlayel}}}				%
\newcommand{\indhidlay}{\ensuremath{l}}							%
\newcommand{\wnn}{\ensuremath{W}}								%
\newcommand{\wnnmat}{\ensuremath{\bo{\wnn}}}					%
\newcommand{\dista}{\ensuremath{d}} 							%

\newcommand{\semvar}{\ensuremath{s}} 							%
\newcommand{\semvarreal}{\ensuremath{\mathsf{\semvar}}} 		%
\newcommand{\semvec}{\ensuremath{\bo{\semvar}}} 				%
\newcommand{\rcind}[1][]{\ensuremath{\ifthenelse{\equal{#1}{}}{R_{\mathtxt{C}}}{R_{\mathtxt{C},#1}}}} %
\newcommand{\MI}{\ensuremath{I}} 								%
\newcommand{\mi}[2][]{\ensuremath{\MI_{#1}\lpa#2\rpa}} 			%
\newcommand{\Lossf}{\ensuremath{\mathcal{L}}} 					%
\newcommand{\relvar}{\ensuremath{z}} 							%
\newcommand{\relvarreal}{\ensuremath{\mathsf{z}}} 				%
\newcommand{\relvec}{\ensuremath{\bo{\relvar}}} 				%
\newcommand{\ntx}{\ensuremath{N_{\mathtxt{Tx}}}} 				%
\newcommand{\nrx}{\ensuremath{N_{\mathtxt{w}}}} 				%
\newcommand{\Nrx}{\ensuremath{N_{\mathtxt{Rx}}}} 				%
\newcommand{\Nfeat}{\ensuremath{N_{\mathtxt{Feat}}}} 			%
\newcommand{\Nsemvec}{\ensuremath{N_{\semvar}}} 				%
\newcommand{\Nrelvec}{\ensuremath{N_{\relvar}}} 				%
\newcommand{\setsym}{\ensuremath{M}} 							%
\newcommand{\dom}{\ensuremath{\mathcal{\setsym}}} 				%
\newcommand{\semset}{\ensuremath{\dom_{\semvar}}} 				%
\newcommand{\relset}{\ensuremath{\dom_{\relvar}}} 				%
\newcommand{\xset}{\ensuremath{\dom_{\xvar}}} 					%
\newcommand{\yset}{\ensuremath{\dom_{\yvar}}} 					%
\newcommand{\snrtrain}{\ensuremath{\mathtxt{SNR}_{\mathtxt{train}}}} 	%
\newcommand{\sinfoni}{SINFONY} 									%
\newcommand{\txrx}{Tx/Rx} 					                    %

\newcommand{\txpars}{\ensuremath{\params}} 						%
\newcommand{\Ntxpars}{\ensuremath{N_{\txpars}}} 				%
\newcommand{\Nrxpars}{\ensuremath{N_{\rxpars}}} 				%
\newcommand{\rxpar}{\ensuremath{\varphi}} 						%
\newcommand{\rxpars}{\ensuremath{\bog{\rxpar}}} 				%
\newcommand{\funcrtrick}{\ensuremath{f}} 						%
\newcommand{\ceparams}{\ensuremath{\Lossf_{\txpars, \rxpars}^{\mathtxt{CE}}}}	%
\newcommand{\lr}{\ensuremath{\epsilon}}                                 %
\newcommand{\dataset}[1][]{\ensuremath{\mathcal{D}_{#1}}}                 %
\newcommand{\trainingset}{\ensuremath{\dataset[\mathtxt{T}]}}             %
\newcommand{\sinfonirl}{RL-\sinfoni}							        %

\newcommand{\hdecvar}{\ensuremath{\nu}} 							%
\newcommand{\hdecvec}{\ensuremath{\bog{\hdecvar}}}                  %
\newcommand{\hrelvec}{\ensuremath{\hat{\relvec}}} 		            %
\newcommand{\hrelvectech}{\ensuremath{\tilde{\relvec}}} 		    %
\newcommand{\similarity}[2]{\ensuremath{\mathtxt{sim}\negthinspace\lpa #1,#2\rpa}} %
\newcommand{\simconst}{\ensuremath{\gamma}} 				        %
\newcommand{\simweight}{\ensuremath{w}} 				            %
\newcommand{\simweightvec}{\ensuremath{\bo{\simweight}}} 			%
\newcommand{\Nfinalfeat}{\ensuremath{N_{\modelfeatures}}}			%
\newcommand{\acc}{\ensuremath{\mathcal{A}}}			                %
\newcommand{\relvarvalue}{\ensuremath{\indc}}                       %
\newcommand{\relvarvaluereal}{\ensuremath{\mathsf{\indc}}}          %
\newcommand{\datasetexemplar}{\ensuremath{\dataset[\mathtxt{HK}]}}  %
\newcommand{\relvecreal}{\ensuremath{\bog{\relvarreal}}} 		    %
\newcommand{\bandwidth}{\ensuremath{B}}                             %
\newcommand{\hrelvectechreal}{\ensuremath{\tilde{\relvecreal}}}     %
\newcommand{\iversonbracket}[1][\cdot]{\ensuremath{\left[#1\right]}}            %
\newcommand{\paramsgcm}{\ensuremath{\viparams_{\mathtxt{G}}}} 	    %
\newcommand{\paramspres}{\ensuremath{\params_{\mathtxt{P}}}} 	    %
\newcommand{\centralimagecommhuman}{Classic digital}                %
\DeclareSymbolFont{sfletters}{OML}{cmbrm}{m}{it}                    %
\DeclareSymbolFont{bfsfletters}{OML}{cmbrm}{b}{it}                  %
\DeclareMathSymbol{\snu}{\mathord}{sfletters}{"17}                  %
\DeclareMathSymbol{\bsnu}{\mathord}{bfsfletters}{"17}               %
\newcommand{\hdecvarreal}{\ensuremath{\snu}}                        %
\newcommand{\hdecvecreal}{\ensuremath{\bsnu}}                       %
\newcommand{\integerset}{\ensuremath{\mathcal{I}}}                  %
\newcommand{\pixelset}{\ensuremath{\integerset}}                    %
\newcommand{\importance}{\ensuremath{\iota}}			            %
\newcommand{\importancevec}{\ensuremath{\bog{\importance}}}			%
\newcommand{\modelfeaturesvec}{\ensuremath{\bog{\modelfeatures}}}	%
\newcommand{\permutationel}{\ensuremath{\pi}}	                    %
\newcommand{\selectedset}{\ensuremath{\integerset}}                 %
\newcommand{\Nselectedset}{\ensuremath{\selectedset_{\Nfinalfeat}}} %
\newcommand{\numberlastlayerfeatures}{\ensuremath{\Nh}}
\newcommand{\Nh}{\ensuremath{N_{\hidlayel}}}					    %
\newcommand{\distfunc}[1]{\ensuremath{\dista\left(#1\right)}} 		%
\newcommand{\simweightstressvec}{\ensuremath{\tilde{\bo{\simweight}}}} 	%
\newcommand{\processingcapacity}{\ensuremath{\norm{\simweightstressvec}_{0}}} %
\newcommand{\constvec}{\ensuremath{\bo{\const}}}                    %
\newcommand{\paramsoverall}{\ensuremath{\params_{\mathtxt{O}}}} 	%
\newcommand{\prescat}{$\hdecvec$-categorical} 	                    %
\newcommand{\presfeat}{$\hdecvec$-$\Nfinalfeat$}                    %
\newcommand{\presfeatbog}{$\hdecvec$-$\bog{\Nfinalfeat}$}           %
\newcommand{\presfeatN}[1]{$\hdecvec$-$\Nfinalfeat=#1$}             %
\newcommand{\presjointtraining}{E2E training}                       %
\newcommand{\memorysizelabel}{Knowledge base size}                  %

\newcommand{\abs}[1][\cdot]{\ensuremath{\left|#1\right|}}               %
\newcommand{\abstxt}[1][\cdot]{\ensuremath{|#1|}}                       %
\newcommand{\norm}[1]{\ensuremath{\left\|#1\right\|}}                   %
\newcommand{\trapo}{\ensuremath{T}}                                     %
\newcommand{\realnum}{\ensuremath{\mathbb{R}}}                          %
\newcommand{\ones}{\ensuremath{\bo{1}}}                                 %
\newcommand{\zero}{\ensuremath{\bo{0}}}                                 %

\newcommand{\idxC}{\ensuremath{k}} %

\newcommand{\funcvar}{\ensuremath{f}}                %
\newcommand{\func}[2][]{\ensuremath{\funcvar_{#1}\negthinspace\lpa#2\rpa}} %
\newcommand{\mF}[1][\nC]{\ensuremath{\mathbf{F}_{\nC}}} %

\newcommand{\eA}{\ensuremath{a}} %
\newcommand{\setA}[1][]{\ensuremath{\mathcal{\MakeUppercase{\eA}}_{#1}}} %
\newcommand{\defD}[1][0]{\ensuremath{\lcb\vdwir[\idxC]\,\vert\,\sd_{\idxC,\idxA}\in\ifthenelse{#1>0}{\setA[\idxC,\idxA]}{\setA}\quad\forall\,\idxA=1,\dotsc,2\nA\rcb}} %

\newcommand{\nA}{\ensuremath{N_{\mathtxt{L}}}} %
\newcommand{\nC}{\ensuremath{N_{\mathtxt{C}}}} %

\newcommand{\vrc}[1][]{\ensuremath{\ifthenelse{\equal{#1}{}}{\mathbf{r}_{\mathtxt{C}}}{\mathbf{r}_{\mathtxt{C},#1}}}} %
\newcommand{\rce}[1][]{\ensuremath{\ifthenelse{\equal{#1}{}}{\hat{R}_{\mathtxt{C}}}{\hat{R}_{\mathtxt{C},#1}}}} %

\newcommand{\drc}[1][]{\ensuremath{\ifthenelse{\equal{#1}{}}{R'_{\mathtxt{C}}}{R'_{\mathtxt{C},#1}}}} %

\newcommand{\sd}{\ensuremath{d}} %

\usepackage{pgfplots}
\pgfplotsset{compat=newest}
\usetikzlibrary{pgfplots.groupplots}
\usetikzlibrary{calc,shapes.geometric,shapes.misc,arrows,matrix}
\usepgfplotslibrary{fillbetween}
\usetikzlibrary{shadows}
\usetikzlibrary{fit}
\usetikzlibrary{backgrounds}
\usetikzlibrary{math}

\newlength{\tikzlinewidth}
\newlength{\tikzlinewidththin}
\newcommand{\fontsizetikz}{\small}						%
\newcommand{\fontsizetikzsmall}{\footnotesize}			%
\newcommand{\fontsizeplot}{\small}						%
\newcommand{\fontsizeplotsmall}{\footnotesize}			%

\tikzset{
pic_style/.style={
tanh/.pic={
        \draw[scale=1, domain=-2:2, smooth, variable=\x, draw=blue, fill=white, line width=\tikzlinewidththin]
        plot ({\x}, {2 * (1 / (1+exp(-4*\x))) - 1});
    },
nn/.pic={
\tikzstyle{every pin edge}=[<-,shorten <=1pt]
\tikzstyle{neuron}=[circle, draw = black, minimum size=10pt, inner sep=0pt]
\tikzstyle{input neuron}=[neuron] %
\tikzstyle{output neuron}=[neuron] %
\tikzstyle{hidden neuron}=[neuron] %
\tikzstyle{annot} = [text width=4em, text centered]
\foreach \name / \y in {1,...,3}
\node[input neuron, pin=left:] (I-\name) at (-1, -\y*0.5 + 1) {};
\foreach \name / \y in {1,...,4}
\path[yshift=0.5cm]
node[hidden neuron] (H-\name) at (0,-\y*0.5 + 0.75) {};
\foreach \name / \y in {1,...,3}
\node[output neuron,pin={[pin edge={->}]right:}, right of=H-\name] (O-\name) at (0,-\y*0.5  + 1) {};
\foreach \source in {1,...,3}
\foreach \dest in {1,...,4}
\path (I-\source) edge (H-\dest);
\foreach \source in {1,...,4}
\foreach \dest in {1,...,3}
\path (H-\source) edge (O-\dest);
},
}
}

\newcommand{\observationcamera}[4]{
    \begin{scope}[shift={#1}, rotate=#2, scale=#3, thick, #4]
        \draw[fill=#4!40, rounded corners=2pt] (0,-0.5) rectangle (1,0.5);
        \draw[fill=#4!20] (1,0) ellipse (0.125 and 0.3);
        \draw[dashed] (1,0) -- (2,0.5);
        \draw[dashed] (1,0) -- (2,-0.5);
        \fill[#4!20, opacity=0.5] (1,0) -- (2,0.5) -- (2,-0.5) -- cycle;
    \end{scope}
}

\newcommand{\transceiver}[4]{
    \begin{scope}[shift={#1}, rotate=#2, scale=#3, thick, #4]
        \draw[fill=#4!40, rounded corners=2pt] (0,-0.5) rectangle (1,0.5);
    \end{scope}
}

\ifx\pdfcompliancelevel\pdfdefault%
    \tikzset{
        shadow_style/.style={drop shadow={shadow xshift=0.75mm, shadow yshift=-0.75mm}},
    }
    \newcommand{\opacitylegend}{0.8}
    \newcommand{\opacityfillbetween}{0.3}
\else
    \tikzset{
        shadow_style/.style={},
    }
    \newcommand{\opacitylegend}{1}
    \newcommand{\opacityfillbetween}{1}
\fi

\tikzset{
thesis_picture_style/.style={
outer xsep = 0pt, outer ysep = 0pt, %
>=latex,
urec/.style = {rectangle, line width = \tikzlinewidth, draw = black, fill = white, minimum width = 50.0, minimum height = 20.00, align=center},
urec_round/.style = {urec, rounded corners = 5},
normalrec/.style={rectangle, line width = \tikzlinewidththin, draw = black, fill = none},
urec_thin/.style = {urec, line width = \tikzlinewidththin},
rec_shadow/.style = {rectangle, line width = \tikzlinewidththin, draw = black, fill = white, minimum width = 50.0, minimum height = 20.00, shadow_style},
gradientblock/.style={normalrec, minimum width=115, minimum height=50, rounded corners=10, line width=0.60mm},
invrec/.style = {rectangle, draw = none, fill = none},
arrowthick/.style={->,line width=\tikzlinewidth},
arrowthin/.style={->,line width=\tikzlinewidththin},
linethin/.style={-,line width=\tikzlinewidththin},
linethick/.style={-,line width=\tikzlinewidth},
ucircle/.style = {circle, line width = \tikzlinewidththin, inner sep=0pt, minimum size = 3.50mm, draw = black, fill = none},
linesplit/.style={circle, draw=black, fill=black, inner sep=0pt, minimum height=3},
dashedbox/.style={dashed, dash pattern=on 2.00mm off 1.00mm},
dashedbox2/.style={dashed, dash pattern={on 7pt off 3pt}, rounded corners=10},
arrowthickdashed/.style={arrowthick, dashedbox},
->-/.style={decoration={markings, mark=at position 0.60 with {\arrow[black,line width=\tikzlinewidththin/2]{latex}}}, postaction={decorate}},
arrowthinmiddle/.style={->-,draw=black,line width=\tikzlinewidththin},
hexagon/.style={regular polygon,regular polygon sides=6,rounded corners=0.5mm,line width=\tikzlinewidththin,draw=red,fill=white,minimum width=48.5mm,minimum height=48.5mm},
triangle/.style={regular polygon,shape border rotate=180,regular polygon sides=3,line width=\tikzlinewidththin,draw=black},
nonlin/.style={rectangle,line width=\tikzlinewidththin,draw=black,fill=none, rounded corners=3},
font=\fontsizetikz,
pic_style,
brace_style/.style={decorate, decoration={brace, mirror, amplitude = 6pt}, line width = \tikzlinewidththin},
zigzag_style/.style={decorate, decoration = {zigzag, segment length = 3mm, amplitude = 0.7mm}, line width = \tikzlinewidththin},
},
thesis_picture_style,
}

\newcommand{\scalecmdnet}{0.8}									%

\tikzset{
    cmdlayer/.style={
            thesis_picture_style,
            scale=0.7/\scalecmdnet,
        },
}

\tikzset{
    cmdextension/.style={
            thesis_picture_style,
            scale=0.7*0.8/\scalecmdnet,
            arrowthick/.style={->,line width=0.60mm},
            linethick/.style={-,line width=0.60mm},
            ucircle/.style={circle,line width=0.30mm,draw=black,fill=none, inner sep=0pt, minimum height=10, minimum size = 3.50mm/\scalecmdnet},
        },
}

\tikzset{
    rlsinfony/.style={
            thesis_picture_style,
            inner xsep = 0pt, inner ysep = 0pt, %
            urec/.style = {rectangle, line width = 0.60mm, draw = black, fill = white, minimum width = 50.0, minimum height = 20.00, align=center},
            arrowthick/.style = {->, line width = 0.60mm},
            arrowthickdashed/.style={arrowthick, dashedbox},
            linethick/.style = {-, line width = 0.60mm},
            font = \fontsizetikzsmall,
        },
}

\tikzset{
    sinfony/.style={
            thesis_picture_style,
            inner xsep = 0pt, inner ysep = 0pt, %
        },
}

\tikzset{
    thesis_overview/.style={
            thesis_picture_style,
        },
}

\definecolor{darkgreen}{rgb}{0,0.498039,0}
\definecolor{darkyellow}{rgb}{0.75,0.75,0}
\definecolor{magenta}{rgb}{0.75,0,0.75}
\definecolor{mediumgreen}{rgb}{0.3,0.8,0.5}
\definecolor{faintgreen}{rgb}{0.85,1.0,0.90}

\definecolor{UBBlueDark}{RGB}{28,53,107}
\definecolor{UBBlueMed}{RGB}{13,104,176}
\definecolor{UBBlueVeryLight}{RGB}{192,209,226} %
\definecolor{UBBlueLight}{RGB}{114,179,223}
\definecolor{UBOrangeDark}{RGB}{247,167,3}
\definecolor{UBOrangeLight}{RGB}{255,232,177}
\definecolor{UBRedMed}{RGB}{213,17,48}
\definecolor{UBRedDark}{RGB}{135,39,70}

\colorlet{color_linebox}{black}
\colorlet{color_new}{mediumgreen}
\colorlet{color_besides}{faintgreen}
\colorlet{color_old}{lightgray}
\colorlet{color_benefit}{darkgreen}
\colorlet{color_drawback}{red}

\colorlet{color_cmdnet_trainable}{red}
\colorlet{color_cmdnet_trainable2}{red}
\colorlet{color_cmdnet_hyperparameter}{blue}

\colorlet{color_communications}{gray!20}
\colorlet{color_receiver}{purple}

\colorlet{color_semantic}{blue}
\colorlet{color_encoder}{black}
\colorlet{color_encoder_rl}{darkgreen}
\colorlet{color_classic_encoder}{darkgreen} %
\colorlet{color_channel}{gray}
\colorlet{color_encoderchannel}{darkgreen}
\colorlet{color_encoderchannel_rl}{orange}
\colorlet{color_decoder}{red}
\colorlet{color_resnet}{darkgreen}
\colorlet{color_transmitter}{purple}
\colorlet{color_semantic_channel}{blue}			%
\colorlet{color_semantic_channel1}{blue!70}
\colorlet{color_semantic_channel2}{teal}
\colorlet{color_semantic_channel3}{cyan}
\colorlet{color_semantic_channel4}{lightgray}
\colorlet{color_optimizer}{black}

\colorlet{color_levela}{blue}
\colorlet{color_levelb}{darkgreen}
\colorlet{color_levelc}{orange}
\colorlet{color_hard_decision}{purple}

\definecolor{hawaii1}{HTML}{8C0273}
\definecolor{hawaii2}{HTML}{922A59}
\definecolor{hawaii3}{HTML}{964742}
\definecolor{hawaii4}{HTML}{996330}
\definecolor{hawaii5}{HTML}{9D831E}
\definecolor{hawaii6}{HTML}{97A92A}
\definecolor{hawaii7}{HTML}{80C55F}
\definecolor{hawaii8}{HTML}{66D89C}
\definecolor{hawaii9}{HTML}{6CEBDB}
\definecolor{hawaii10}{HTML}{B3F2FD}

\definecolor{turquoise}{rgb}{0.251, 0.878, 0.816}

\definecolor{qualitative1}{HTML}{e41a1c}    %
\definecolor{qualitative2}{HTML}{377eb8}    %
\definecolor{qualitative3}{HTML}{4daf4a}    %
\definecolor{qualitative4}{HTML}{984ea3}    %
\definecolor{qualitative5}{HTML}{ff7f00}    %
\definecolor{qualitative6}{HTML}{ffff33}    %

\colorlet{color_communications_design}{darkgreen}
\colorlet{color_human_decision_making}{red}
\colorlet{color_human_machine_interface}{magenta}
\colorlet{color_sinfony_decision_making}{darkgray}                   %
\colorlet{color_human_decision_making_features_1}{red}               %
\colorlet{color_human_decision_making_features_max}{turquoise}       %
\colorlet{color_human_decision_making_features_40}{blue}             %
\colorlet{color_human_decision_making_features_20}{orange}           %
\colorlet{color_human_decision_making_features_10}{qualitative3}     %
\colorlet{color_human_decision_making_features_5}{qualitative4}      %

\colorlet{color_sd}{darkgray}
\colorlet{color_mmse}{darkgray}
\colorlet{color_mosic}{purple} 			%
\colorlet{color_amp}{darkgreen} 		%
\colorlet{color_sdr}{orange} 			%
\colorlet{color_mmnet}{darkyellow} 		%
\colorlet{color_detnet}{magenta} 		%
\colorlet{color_oamp}{blue} 			%
\colorlet{color_cmd}{red} 				%
\colorlet{color_cmdshallow}{orange} 	%
\colorlet{color_hypercmd}{darkgreen} 	%
\colorlet{color_cmdpar}{darkgreen} 		%
\colorlet{color_awgn}{green!50!black} 	%
\colorlet{color_mf}{black} 				%
\colorlet{color_mfvi_sic}{red} 			%
\colorlet{color_mfvi_pic}{blue} 		%

\colorlet{color_cmd_default}{purple}
\colorlet{color_cmd_splin}{color_cmd}
\colorlet{color_cmd_splin_iteration}{blue}
\colorlet{color_cmd_default_fixed}{darkgreen}
\colorlet{color_cmd_splin_fixed}{darkgreen}
\colorlet{color_cmd_ldpc}{red}
\colorlet{color_cmd_NL4}{purple} 		%
\colorlet{color_cmd_NL8}{blue} 			%
\colorlet{color_cmd_NL32}{darkgray} 	%
\colorlet{color_cmd_NL128}{brown} 		%
\colorlet{color_parameters}{red}
\colorlet{color_parameters_starting}{blue}
\colorlet{color_mse_loss}{orange}
\colorlet{color_multi_ce_loss}{darkgreen}
\colorlet{color_optimizer_sgd}{blue}
\colorlet{color_cmd_online}{color_cmd}
\colorlet{color_dnn_online}{color_detnet}
\colorlet{color_minima}{red}
\colorlet{color_objective}{blue}

\colorlet{concrete1}{red}
\colorlet{concrete2}{darkgray}
\colorlet{concrete3}{orange}
\colorlet{concrete4}{darkgreen}
\colorlet{concrete5}{blue}

\definecolor{color0}{rgb}{0.12156862745098,0.466666666666667,0.705882352941177}
\definecolor{color1}{rgb}{1,0.498039215686275,0.0549019607843137}
\definecolor{color2}{rgb}{0.172549019607843,0.627450980392157,0.172549019607843}
\definecolor{color3}{rgb}{0.83921568627451,0.152941176470588,0.156862745098039}

\colorlet{color_mean}{color_sd}
\colorlet{color_map}{color_sdr}
\colorlet{color_single_bit}{color_oamp}
\colorlet{color_dnn_estimator}{color_cmd}
\colorlet{color_analog_transmission}{color_amp}
\colorlet{color_dnn_transceiver}{purple}

\colorlet{color_central_classification}{color_sd}
\colorlet{color_sinfony_perfect}{color_oamp}
\colorlet{color_sinfony_perfect_awgn}{color_oamp}
\colorlet{color_sinfony_perfect_training_awgn}{color_oamp}
\colorlet{color_sinfony_transceiver_ntx64}{color_cmd}
\colorlet{color_sinfony_transceiver_ntx16}{color_amp}
\colorlet{color_image_digital_communication}{color_sd}
\colorlet{color_feature_digital_communication}{color_sd}
\colorlet{color_feature_analog_communication}{color_detnet}
\colorlet{color_rlsinfony}{color_oamp}
\colorlet{color_sinfony_perfect_rlpaper}{darkgreen}
\colorlet{color_sinfony_transceiver_ntx16_rlpaper}{color_cmd}

\colorlet{color_sinfony_rx}{purple}
\colorlet{color_sinfony_layer}{orange}

\newlength{\smalllinewidth}
\newlength{\largelinewidth}
\newlength{\grouplabelshift}
\newlength{\subgrouplabelshift}
\pgfplotsset{
    marker_circle/.style={mark=*, mark size=\largelinewidth, mark options={solid}},
    marker_square/.style={mark=square*, mark size=\largelinewidth, mark options={solid}},
    marker_x/.style={mark=x, mark size=4*\smalllinewidth, mark options={solid}},
    marker_x_small/.style={marker_x, mark size=3*\smalllinewidth},
    marker_triangle/.style={mark=triangle*, mark size=\largelinewidth, mark options={solid}},
    marker_triangle_large/.style={marker_triangle, mark size=3*\smalllinewidth, mark options={solid}},
    marker_diamond/.style={mark=diamond*, mark size=\largelinewidth, mark options={solid}},
    marker_diamond_large/.style={mark=diamond*, mark size=3*\smalllinewidth, mark options={solid}},
    marker_star/.style={mark = star, mark size = 3.5*\smalllinewidth, mark options={solid}},
    marker_asterisk/.style={mark=asterisk, mark size=3.5*\smalllinewidth, mark options={solid}},
    marker_plus/.style={mark=+, mark size=3.5*\smalllinewidth, mark options={solid}},
}
\pgfplotsset{
    pt_concrete1/.style={color=concrete1, marker_circle},
    pt_concrete2/.style={color=concrete2, marker_triangle, mark size=2.3*\smalllinewidth},
    pt_concrete3/.style={color=concrete3, marker_square, dashed},
    pt_concrete4/.style={color=concrete4, marker_square},
    pt_concrete5/.style={color=concrete5, marker_diamond_large},
    pt_concrete6/.style={pt_concrete4, marker_circle, dashed},
    ptsd/.style={color=color_sd, marker_circle}, 						%
    ptmmse/.style={color=color_mmse, marker_triangle, dashed}, 			%
    ptmosic/.style={color=color_mosic, marker_triangle}, 				%
    ptamp/.style={color=color_amp, marker_diamond}, 					%
    ptsdr/.style={color=color_sdr, marker_circle, dashed},				%
    ptdetnet/.style={color=color_detnet, marker_square, dashed},		%
    ptmmnet/.style={color=color_mmnet, marker_diamond, dashed},			%
    ptoamp/.style={color=color_oamp, marker_x},							%
    ptcmd/.style={color=color_cmd, marker_square},						%
    ptcmdshal/.style={color=color_cmdshallow, marker_square, dashed},	%
    ptawgn/.style={color=color_awgn, semithick}, 						%
    ptcmd_default/.style={color=color_cmd_default, marker_diamond_large}, 				%
    ptcmd_splin/.style={color=color_cmd_splin, marker_triangle_large, dashed}, 			%
    ptcmd_default_fixed/.style={color=color_cmd_default_fixed, marker_x_small, dashed}, %
    ptcmd_splin_fixed/.style={color=color_cmd_splin_fixed, marker_x_small}, 			%
    ptcmd_ldpc/.style={color=color_cmd_ldpc, marker_circle}, 			%
    ptamp_NL8/.style={ptamp, marker_star},								%
    ptamp_NL16/.style={ptamp, dashed},									%
    ptcmd_NL4/.style={color=color_cmd_NL4, marker_x},					%
    ptcmd_NL8/.style={color=color_cmd_NL8, marker_triangle},			%
    ptcmd_NL32/.style={color=color_cmd_NL32, marker_triangle, dashed},	%
    ptcmd_NL128/.style={color=color_cmd_NL128, marker_x, dashed},		%
    pthcmd/.style={color=color_hypercmd},								%
    ptcmdpar/.style={color=color_cmdpar},								%
    ptcmd_mismatch/.style={ptcmd, marker_asterisk, dashed},
    ptcmd_online_Ne10e2/.style={color=color_cmd_online, marker_diamond, dashed},
    ptdnn_online_Ne10e2/.style={color=color_dnn_online, marker_star, dashed},
    ptdnn_online_Ne10e3/.style={color=color_dnn_online, marker_plus, dashdotted},
    ptdnn_online_Ne10e4/.style={color=color_dnn_online, marker_x, dotted},
    ptdnn_online_Ne10e5/.style={color=color_dnn_online, marker_star, densely dotted},
    pt_parameters/.style={color=color_parameters},
    pt_parameters_starting/.style={color=color_parameters_starting, dashed},
    pt_histogram/.style={ybar legend, line width=0.01pt}, %
    pt_barplot/.style={ybar legend, bar width=0.75, draw=none},
    pt_3dplot/.style={surf, faceted color = color_objective!60, fill = color_objective!20},
    pt_3dplot_points/.style={only marks, draw = color_minima, fill = color_minima!60, marker_circle},
    pt_2dplot/.style={blue},
    pt_training/.style={semithick},
    pt_mse_loss/.style={color_mse_loss, marker_circle, dashed},
    pt_multi_ce_loss/.style={color_multi_ce_loss, marker_triangle, mark options={solid, rotate=270}, dashed},
    pt_optimizer_adam_Nb32/.style={ptcmd, marker_x},
    pt_optimizer_adam_Nb5000/.style={ptcmd, mark=None},
    pt_optimizer_sgd/.style={ptcmd, color_optimizer_sgd, dashed},
    pt_optimizer_sgd_Nb32/.style={pt_optimizer_adam_Nb32, color_optimizer_sgd, dashed},
    pt_optimizer_sgd_Nb5000/.style={pt_optimizer_adam_Nb5000, color_optimizer_sgd, dashed},
    pt_mfvi_sic/.style={color=color_mfvi_sic, marker_star}, 			%
    pt_mfvi_pic/.style={color=color_mfvi_pic, marker_plus}, 			%
}
\pgfplotsset{
    plot_mean/.style={ptsd, color=color_mean},
    plot_map/.style={ptsdr, color=color_map},
    plot_single_bit/.style={ptoamp, color=color_single_bit},
    plot_dnn_estimator/.style={ptcmd, color=color_dnn_estimator},
    plot_analog_transmission/.style={ptamp, color=color_analog_transmission},
    pt_central_classification/.style={color=color_central_classification, dashed, mark = None},
    pt_sinfony_perfect/.style={color=color_sinfony_perfect, dashed, mark = None},
    pt_sinfony_perfect_awgn/.style={color=color_sinfony_perfect_awgn, mark = None},
    pt_sinfony_perfect_training_awgn/.style={ptoamp,color=color_sinfony_perfect_training_awgn},
    pt_sinfony_transceiver_ntx64/.style={ptcmd,color=color_sinfony_transceiver_ntx64},
    pt_sinfony_transceiver_ntx16/.style={ptamp,color=color_sinfony_transceiver_ntx16},
    pt_sinfony_transceiver_ntx8/.style={color=color_sinfony_transceiver_ntx16, marker_x, dashed},
    pt_sinfony_transceiver_ntx4/.style={color=color_sinfony_transceiver_ntx16, marker_circle, dashed},
    pt_sinfony_transceiver_ntx2/.style={color=color_sinfony_transceiver_ntx16, marker_triangle_large, dashed},
    pt_image_digital_communication/.style={color=color_image_digital_communication, marker_x},
    pt_feature_digital_communication1/.style={ptsd, color=color_feature_digital_communication},
    pt_feature_digital_communication2/.style={color=color_feature_digital_communication, marker_diamond, dashed},
    pt_feature_digital_communication3/.style={color=color_feature_digital_communication, marker_star, dashed},
    pt_feature_analog_communication/.style={color=color_feature_analog_communication, marker_triangle, dashed},
    pt_rlsinfony_transceiver_ntx16/.style={color=color_rlsinfony, marker_x}, %
    pt_rlsinfony_transceiver_ntx16_2/.style={pt_rlsinfony_transceiver_ntx16, marker_triangle, dotted}, %
    pt_sinfony_perfect_rlpaper/.style={pt_sinfony_perfect, color=color_sinfony_perfect_rlpaper},
    pt_sinfony_transceiver_ntx16_rlpaper/.style={pt_sinfony_transceiver_ntx64, color=color_sinfony_transceiver_ntx16_rlpaper}, %
    pt_sinfony_transceiver_ntx16_rlpaper_2/.style={pt_sinfony_transceiver_ntx16_rlpaper, dotted}, %
    pt_sinfony_convergence/.style={color=color_sinfony_transceiver_ntx16_rlpaper, mark = None, line width = 0.7pt, opacity = \opacityfillbetween},
    pt_rlsinfony_convergence/.style={pt_sinfony_convergence, color=color_rlsinfony},
    pt_sinfony_convergence_fill/.style={pt_sinfony_convergence, color=color_sinfony_transceiver_ntx16_rlpaper!30!white},
    pt_rlsinfony_convergence_fill/.style={pt_rlsinfony_convergence, color=color_rlsinfony!30!white},
    pt_sinfony_convergence_border/.style={pt_sinfony_convergence, color=color_sinfony_transceiver_ntx16_rlpaper!50!white}, %
    pt_rlsinfony_convergence_border/.style={pt_rlsinfony_convergence, color=color_rlsinfony!50!white}, %
    pt_sinfony_transceiver_ntx16_snr/.style={pt_sinfony_transceiver_ntx16, dotted},
    pt_sinfony_individual_rx/.style={pt_sinfony_transceiver_ntx16, color=color_sinfony_rx, marker_x},
    pt_sinfony_rxjoint/.style={pt_sinfony_individual_rx, dashed},
    pt_sinfony_txrx_layer2/.style={pt_sinfony_transceiver_ntx16, color=color_sinfony_layer, marker_asterisk},
    pt_sinfony_txrx_layer3/.style={pt_sinfony_txrx_layer2, dashed},
    pt_hdm_features_1/.style={
            color=color_human_decision_making_features_1,
            semithick,
        },
    pt_hdm_features_5/.style={
            color=color_human_decision_making_features_5,
            semithick,
        },
    pt_hdm_features_10/.style={
            color=color_human_decision_making_features_10,
            semithick,
        },
    pt_hdm_features_20/.style={
            color=color_human_decision_making_features_20,
            semithick,
        },
    pt_hdm_features_40/.style={
            color=color_human_decision_making_features_40,
            semithick,
        },
    pt_hdm_features_max/.style={
            color=color_human_decision_making_features_max,
            semithick,
        },
    pt_hdm_sinfony/.style={
            color=color_sinfony_decision_making,
            semithick,
            dashed,
        },
    line_joint_training/.style={
            dash dot,
        },
    pt_hdm_features_1_joint/.style={
            color=color_human_decision_making_features_1,
            semithick,
            line_joint_training,
        },
    pt_hdm_features_5_joint/.style={
            color=color_human_decision_making_features_5,
            semithick,
            line_joint_training,
        },
    pt_hdm_features_10_joint/.style={
            color=color_human_decision_making_features_10,
            semithick,
            line_joint_training,
        },
    pt_hdm_features_20_joint/.style={
            color=color_human_decision_making_features_20,
            semithick,
            line_joint_training,
        },
    pt_hdm_features_40_joint/.style={
            color=color_human_decision_making_features_40,
            semithick,
            line_joint_training,
        },
    pt_hdm_features_max_joint/.style={
            color=color_human_decision_making_features_max,
            semithick,
            line_joint_training,
        },
}

\pgfplotsset{
    plotsize/.style={
            height = 7cm, %
            width = 0.95\textwidth, %
            group style={
                    group size=2 by 1,
                    horizontal sep=1.25cm, %
                    every plot/.style= {
                            width=0.52\textwidth, %
                            height=6.5cm, %
                        },
                },
        },
    plotsize_onlyaxis/.style={
            scale only axis, 			%
            height = 5.5cm, %
            width = 9cm,	%
            group style={
                    group size=2 by 1,
                    every plot/.style= {
                            scale only axis,
                            width=4.75cm, 	%
                            height=5cm, 	%
                        },
                },
        },
}

\pgfplotsset{
    thesis_axis_style/.style={
            plotsize,
            every axis plot/.style={line width=\largelinewidth},
            tick pos=left,
            tick align=inside,%
            xminorticks=true, %
            xminorgrids=true, %
            xmajorgrids=true, %
            yminorgrids=true, %
            ymajorgrids=true, %
            zmajorgrids=true, %
            every outer x axis line/.style={line width=\largelinewidth},
            every outer y axis line/.style={line width=\largelinewidth},
            every outer z axis line/.style={line width=\largelinewidth},
            every tick/.style={
                    line width=\smalllinewidth,
                },
            label style={font=\fontsizeplot},
            title style={font=\fontsizeplot},
            every y tick label/.style={font=\fontsizeplot},
            every x tick label/.style={font=\fontsizeplot},
            every z tick label/.style={font=\fontsizeplot},
            legend style= {
                    line width=\smalllinewidth,
                    font=\fontsizeplotsmall,
                    legend cell align=left,
                    align=left,
                    fill=white,
                    fill opacity=\opacitylegend, draw opacity=1, text opacity=1,
                },
            /pgfplots/layers/barplotaxis/.define layer set={
                    axis background,axis grid, pre main,main,axis ticks,axis lines,axis tick labels,axis descriptions,axis foreground
                }{/pgfplots/layers/standard},
            /pgfplots/layers/rlplotaxis/.define layer set={
                    pre main, axis background,axis grid,main,axis ticks,axis lines,axis tick labels,axis descriptions,axis foreground
                }{/pgfplots/layers/standard},
        },
    thesis_axis_style,
}

\pgfplotsset{
    journal_axis_style/.style={
            every axis plot/.style={line width=1.4pt},
            tick pos=left,
            yminorgrids,
            every outer x axis line/.style={line width=1.4pt},
            every outer y axis line/.style={line width=1.4pt},
            every tick/.style={
                    black,
                    line width=0.7pt,
                },
            label style={font=\fontsize{8}{9}\selectfont},
            title style={font=\fontsize{8}{9}\selectfont},
            every y tick label/.style={font=\fontsize{8}{9}\color{black}},
            every x tick label/.style={font=\fontsize{8}{9}\color{black}},
            every extra x tick/.style={grid style={solid, violet, line width=2.8pt},
                    x tick label style={/pgf/number format/.cd,precision=10}
                },
            legend style= {
                    line width=0.7pt,
                    font=\fontsize{11.4}{12}\selectfont\color{black},
                    legend cell align=left,
                    align=left,
                    fill=white,
                    fill opacity=0.8, draw opacity=1, text opacity=1,
                    nodes={scale=0.7, transform shape},
                },
            height = 8cm, %
        }
}

\usepackage{scalerel}
\usetikzlibrary{svg.path}
\definecolor{orcidlogocol}{HTML}{A6CE39}
\tikzset{
    orcidlogo/.pic={
            \fill[orcidlogocol] svg{M256,128c0,70.7-57.3,128-128,128C57.3,256,0,198.7,0,128C0,57.3,57.3,0,128,0C198.7,0,256,57.3,256,128z};
            \fill[white] svg{M86.3,186.2H70.9V79.1h15.4v48.4V186.2z}
            svg{M108.9,79.1h41.6c39.6,0,57,28.3,57,53.6c0,27.5-21.5,53.6-56.8,53.6h-41.8V79.1z M124.3,172.4h24.5c34.9,0,42.9-26.5,42.9-39.7c0-21.5-13.7-39.7-43.7-39.7h-23.7V172.4z}
            svg{M88.7,56.8c0,5.5-4.5,10.1-10.1,10.1c-5.6,0-10.1-4.6-10.1-10.1c0-5.6,4.5-10.1,10.1-10.1C84.2,46.7,88.7,51.3,88.7,56.8z};
        }
}

\newcommand\orcidicon[1]{\href{https://orcid.org/#1}{\mbox{\scalerel*{
                \begin{tikzpicture}[yscale=-1,transform shape]
                    \pic{orcidlogo};
                \end{tikzpicture}
            }{U}}}} %

\usepackage{TikZ/graphical_abstract/antenna}
\usepackage{TikZ/graphical_abstract/computer}
\usepackage{TikZ/graphical_abstract/person}
\usepackage{TikZ/graphical_abstract/phone}
\usepackage{TikZ/graphical_abstract/satellite}
\usepackage{TikZ/graphical_abstract/satellitedish}
\usepackage{colorprofiles}
\newcommand{\pdfdefault}{a-2b}
\newcommand{\pdfcompliancelevel}{a-2b}    %

\begin{document}

\title{Integrating Semantic Communication and Human Decision-Making into an End-to-End Sensing-Decision Framework}

\author{Edgar Beck$^{\orcidicon{0000-0003-2213-9727}}$,~\IEEEmembership{Graduate Student,~IEEE,}
    Hsuan-Yu Lin$^{\orcidicon{0000-0002-2570-5379}}$,
    Patrick Rückert$^{\orcidicon{0000-0002-9489-592X}}$,
    Yongping Bao$^{\orcidicon{0000-0001-6399-3300}}$,
    Bettina von Helversen$^{\orcidicon{0000-0003-2004-6922}}$,
    Sebastian Fehrler$^{\orcidicon{0000-0002-5220-5943}}$,
    Kirsten Tracht$^{\orcidicon{0000-0001-9740-3962}}$,
    Armin Dekorsy$^{\orcidicon{0000-0002-5790-1470}}$,~\IEEEmembership{Senior,~IEEE}
    \thanks{This work was partly funded by the Federal State of Bremen and the University of Bremen as part of the Humans on Mars Initiative, by the German Ministry of Education and Research (BMBF) under grant 16KISK016 (Open6GHub), and by the German Research Foundation (DFG) under grant 500260669 (SCIL).}
    \thanks{Edgar Beck and Armin Dekorsy are with the Department of Communications Engineering, University of Bremen, 28359 Bremen, Germany (e-mail: \{beck, dekorsy\}@ant.uni-bremen.de). Patrick Rückert and Kirsten Tracht are with the Bremen Institute for Mechanical Engineering, University of Bremen (e-mail: \{rueckert, tracht\}@uni-bremen.de). Hsuan-Yu Lin and Bettina von Helversen are with the Department of Psychology, Faculty of Human and Health Sciences, University of Bremen (e-mail: \{hslin, b.helversen\}@uni-bremen.de). Yongping Bao and Sebastian Fehrler are with the SOCIUM Research Center on Inequality and Social Policy, University of Bremen (e-mail: \{yongping.bao, sebastian.fehrler\}@uni-bremen.de).}
}

\maketitle{}

\begin{abstract}
    As early as 1949, Weaver defined communication in a very broad sense to include all procedures by which one mind or technical system can influence another, thus establishing the idea of semantic communication. With the recent success of machine learning in expert assistance systems where sensed information is wirelessly provided to a human to assist task execution, the need to design effective and efficient communications has become increasingly apparent. In particular, semantic communication aims to convey the meaning behind the sensed information relevant for Human Decision-Making (HDM). Regarding the interplay between semantic communication and HDM, many questions remain, such as how to model the entire end-to-end sensing-decision-making process, how to design semantic communication for the HDM and which information should be provided for HDM\@. To address these questions, we propose to integrate semantic communication and HDM into one probabilistic end-to-end sensing-decision framework that bridges communications and psychology. In our interdisciplinary framework, we model the human through a HDM process, allowing us to explore how feature extraction from semantic communication can best support HDM both in theory and in simulations. In this sense, our study reveals the fundamental design trade-off between maximizing the relevant semantic information and matching the cognitive capabilities of the HDM model. Our initial analysis shows how semantic communication can balance the level of detail with human cognitive capabilities while demanding less bandwidth, power, and latency.
\end{abstract}


\section{Introduction}%
\label{sec:introduction}
\IEEEPARstart{W}{ith} recent breakthroughs in Machine Learning (ML), such as generative Artificial Intelligence (AI) or Natural Language Processing (NLP), assistance systems are now finding their way into everyday life~\cite{vaswani_attention_2017}. For example, doctors are supported by expert assistance systems that outperform human expertise in evaluating medical image data for disease diagnosis~\cite{ciresan_mitosis_2013}. Many assistance systems acquire information about physical, chemical, and biological processes through sensors or sensor networks and transmit it to humans for decision-making when performing specific tasks. Applications that exploit such assistance systems include remote operation concepts such as digital twins for production, rescue scenarios, healthcare, autonomous driving, underwater repairs, remote sensing for earth observation and swarm exploration~\cite{suresh_human-integrated_2024}. For example, mobile robotic systems equipped with sensors can assist Human Decision-Making (HDM). All of this relies heavily on efficient and effective wireless communications, which is therefore an integral part of the entire end-to-end sensing-decision-making process.

At this point, semantic communication comes into play as it deals with the question of how information from the assistance system can be communicated more effectively to the human to improve HDM in task execution while demanding less bandwidth, power, and latency. Several research questions can be identified from this interplay:
\begin{itemize}
    \item[a)] \textbf{Joint Modeling:} How to model the end-to-end sensing-decision-making process that bridges the disciplines communications and psychology?
    \item[b)] \textbf{Suitability \& Design:} Is semantic communication suitable for providing the information needed in terms of relevance and accuracy to facilitate effective HDM\@? Given a task, how to design/optimize semantic communication for accurate HDM, specifically:
          \begin{itemize}
              \item[i)] Which information should be provided to HDM?
              \item[ii)] How much information should be provided to HDM?
          \end{itemize}
    \item[c)] \textbf{Impact:} Given the provided semantic information, how does the HDM process impact the end-to-end sensing-decision-making process?
\end{itemize}
To address these questions, we propose integrating semantic communication and HDM into a unified probabilistic end-to-end sensing-decision framework, thereby composing all three levels described by Weaver~\cite{shannon_mathematical_1949}. To showcase our framework's applicability and highlight its key mechanisms, we examine a case study grounded in an empirical categorization example. As a starting point of our study, we will first reflect upon the state of the art in semantic communication and human decision-making.

\subsection{Semantic Communication}

In the 1949 review of Shannon's general theory of communication~\cite{shannon_mathematical_1949}, Weaver introduces the idea of semantic communication with regard to both humans and technical systems. There, he used the term communication \emph{``in a very broad sense to include all of the procedures by which one mind may affect another. This, of course, involves not only written and oral speech, but also music, the pictorial arts, the theatre, the ballet, and in fact all human behavior. In some connections it may be desirable to use a still broader definition of communication, namely, one which would include the procedures by means of which one mechanism [\ldots] %
    affects another mechanism [\ldots]%
    .''}
To meet the unprecedented demands of 6G communication efficiency in terms of bandwidth, latency, and power, attention has been drawn to the broad concept of semantic communication~\cite{shannon_mathematical_1949, calvanese_strinati_6g_2021,luo_semantic_2022,wheeler_engineering_2023, gunduz_beyond_2023, beck_semantic_2023}, supported by the idea of an 6G AI-native air interface~\cite{hoydis_toward_2021}. It aims to transmit the meaning of a message rather than its exact version, which has been the focus of digital error-free system design~\cite{shannon_mathematical_1949}. Approaches to the description or design of semantic communication can be divided into statistical probability-based~\cite{shao_learning_2022}, logical probability-based~\cite{carnap_outline_1952}, knowledge graph-based~\cite{zhou_cognitive_2024}, and kernel-based~\cite{pokhrel_understand-before-talk_2023}.

Arguing for the generality of Shannon's theory not only for the technical level but for the semantic level design as Weaver~\cite{shannon_mathematical_1949}, Bao, Basu et al.~\cite{bao_towards_2011, basu_preserving_2014} were the first to define semantic information sources and channels to tackle the semantic design by information-theoretic approaches.

With the rise of Machine Learning (ML) in communication research, transformer-based Deep Neural Networks (DNNs) have been introduced to AutoEncoders (AEs) for text transmission to learn compressed hidden representations of semantic content, aiming to improve communication efficiency~\cite{xie_deep_2021}. However, accurate recovery of the source (text) is the main goal. The approach improves performance in semantic metrics, especially at low Signal-to-Noise Ratio (SNR), compared to classical digital transmissions. It has been adapted to many other problems,~e.g., speech transmission~\cite{weng_semantic_2021, weng_deep_2023}. Meanwhile, also recent advances in large AI models have found their way into semantic communication~\cite{jiang_large_2024,cui_llmind_2024}.

From a theoretical perspective, building upon the ideas of Bao, Basu et al.~\cite{bao_towards_2011, basu_preserving_2014}, in~\cite{beck_semantic_2023,beck_model-free_2024}, the authors explicitly define a semantic random variable and identify the Information Maximization (InfoMax) problem and its variation, the Information Bottleneck (IB) problem, as appropriate semantic design criteria. Solving the InfoMax problem with ML tools, the authors obtain their design Semantic INFOrmation TraNsmission and RecoverY (\sinfoni{}). For more details, we refer the reader to \refsec{sec:semantic_communication} and~\refsec{sec:sinfony_numres}. Furthermore, task-oriented edge-cloud transmission has been formulated as an IB problem~\cite{shao_learning_2022}.

Semantic communication has been extended to process several types of data,~i.e., multimodal data, such as image, text, depth map data~\cite{xie_task-oriented_2022,luo_multimodal_2024}. In addition, monitoring, planning, and control of real worlds require the processing of multiple tasks. Thus, in~\cite{halimi_semantic_2024, halimi_cooperative_2024}, the authors extend the concept of a semantic source to include multiple semantic interpretations. To facilitate cooperative multitask processing and improve training convergence, the semantic encoders are divided into common and specific units, extracting common low-level features and separate high-level features. %
In~\cite{wen_task-oriented_2024}, the authors move beyond traditional data-level compression towards a joint sensing, computation, and communication framework that emphasizes sensing, power and timing constraints. They propose a holistic semantic-level system optimization involving sensing quality, feature quantization, and resource constraints, achieved through formulating and solving resource allocation problems directly aimed at improving task performance.

So far, the human behind the application or task has only been taken into account by theory, with the rate-distortion-perception trade-off~\cite{blau_rethinking_2019,chai_rate-distortion-perception_2023}. For example, the mean square error distortion is known to be inconsistent with human perception and thus not a good semantic optimization criterion~\cite{blau_rethinking_2019}. Precisely because humans make the final decision when performing a task, we aim to fill the research gap of bridging semantic communication and human decision-making into an end-to-end sensing-decision framework.

\subsection{Human Decision-Making}

Even though the decision capability of artificial systems is increasing, in many situations the final decision-maker will be a human, and humans do not always make rational decisions. Therefore, to optimize the results, the needs, and capabilities of the decision-maker must be considered in the semantic communication design,~e.g., by definition of the semantic source.

Humans are undoubtedly expert decision-makers who can cope well with uncertainty and complexity~\cite{payne_constructive_1992, hassabis2017neuroscience}. However, it has been repeatedly shown that Human Decision-Making (HDM) can be systematically biased and decisions can be influenced by irrelevant information and context, as shown in the large literature on heuristics and biases~\cite{kahneman_perspective_2013}. For example, judges' sentencing decisions can be systematically influenced by asking whether a sentence should be higher or lower than a randomly generated number~\cite{englich_playing_2006}, and decisions differ depending on whether the same information is presented in frequencies or percentages~\cite{thomas_heuristics_2009}.

Rational models of decision-making typically require the decision-maker to consider all relevant information about the decision options and the context~\cite{gigerenzer_heuristic_2011}. However, humans have limited cognitive resources, such as attention or working memory capacity, which restricts the amount of information they can process at once~\cite{cowan_magical_2001, simon_bounded_1991}.

It is often assumed that humans deal with these limited capacities by using simplified decision strategies that often consider only a subset of the information and discard ``extra'' information~\cite{simon_bounded_1991, kahneman_perspective_2013, gigerenzer_heuristic_2011}. For example, the ``take-the-best'' heuristic assumes that the decision-maker considers only one dimension at a time in the order of validity of the dimension. A decision is made when the decision-maker encounters a dimension that discriminates between alternatives~\cite{gigerenzer_betting_1999}. Importantly, the use of heuristics such as the take-the-best heuristic often leads to decision performance on par with complex decision rules if the most valid predictors are indeed considered~\cite{brighton_robust_2006}.

However, humans are not always able to identify the best predictors, especially when the information environment is complex, they lack expertise, are pressed for time, or are distracted~\cite{gigerenzer_reasoning_1996}. In these situations, as the growing literature on decision-support/assistance systems shows, human decision-making can be supported and improved by highlighting relevant information, providing summary information, or reducing irrelevant information~\cite{paul_input_2010}. Even when the human decision-maker has access to all relevant information and is able to integrate the information properly, humans have a tendency to respond probabilistically~\cite{friedman_monty_1998, sanborn_noise_2024}. When given several options, and each option has a certain probability of being correct, the optimal decision (that has the highest chance of being correct) is to deterministically choose the option with the highest probability of being correct. While humans choose the best option in the majority of the trials, they usually also tend to choose other options. This variability in human decision-making has likely multiple causes~\cite{seitz_disentangling_2024}.

Semantic communication offers the flexibility to adapt the transmitted information to facilitate the achievement of the human decision-maker's goals. The integration of semantic communication and human decision-making leads to a paradigm shift that includes the communication chain in the decision-support/assistance system.

\subsection{Main Contributions}

The main contributions addressing the above-mentioned research questions are the following:
\begin{itemize}
    \item In this article, we propose a broadly applicable probabilistic end-to-end sensing-decision framework that wirelessly links sensed data with relevant information-based Human-Decision Making (HDM) by semantic communication.
    \item Based on this framework, we extend the information-theoretic view on semantic communication towards presentation design and HDM model training, revealing the fundamental presentation or semantic communication design trade-off between maximizing the relevant semantic information and matching the cognitive capabilities of the HDM model. In this sense, our study provides new insights for the design/interaction of semantic communication with models of HDM\@.
    \item To showcase our framework's applicability and investigate its key mechanisms, we examine categorization examples using effective HDM models. Simulation results on image and audio stimuli data show that, when balancing the design trade-off between feature extraction in semantic communication and cognitive constraints of the HDM model, adjusting the level of detail to match human cognitive capabilities is more important for achieving high decision accuracy than simply providing more relevant information. Moreover, this effect becomes more pronounced when the context changes,~e.g., under limited experience or processing capacity. Furthermore, randomness involved in the HDM process decreases accuracy.
    \item Semantic communication is able to provide the HDM model with sufficient information for making accurate decisions, while demanding less bandwidth, power, and latency compared to classical digital Shannon-based approaches.
    \item We exploit our end-to-end framework to jointly optimize semantic communication and HDM. First numerical results show that end-to-end optimization has the potential to improve HDM accuracy at the cost of additional alternating training iterations.
    \item Finally, we provide an outlook on open research questions of our approach, including the design of information presentation through visualization, as well as game theory perspectives on sender-receiver conflicts of interest.
\end{itemize}

\section{End-to-End Sensing-Decision Framework}%
\label{sec:e2e_framework}

\begin{figure}[!t]%
    \centering
    \input{TikZ/e2e_sensing_decision_tool_example.tikz}
    \caption{Sketch of the end-to-end sensing-decision process for the example of tool wear assessment. It also situates the fundamental design trade-off between semantic communication and human decision-making.}%
    \label{fig0:e2e_tool_example}
\end{figure}

\begin{figure}[!t]%
    \centering
    \begin{tikzpicture}[%
		inner xsep = 0pt, inner ysep = 0pt, %
		outer xsep = 0pt, outer ysep = 0pt, %
		> = latex,
		ucircle/.style = {circle, line width = 0.30mm, minimum size = 4.00mm, draw = black, fill = white},
		urec/.style = {rectangle, line width = 0.60mm, draw = black, fill = white, minimum width = 50.0, minimum height = 20.00},
		invrec/.style = {rectangle, line width = 0.30mm, draw = none, fill = none, minimum width = 50.0, minimum height = 20.00},
		arrowthick/.style = {->, line width = 0.6mm},
		arrowthin/.style = {<->, line width = 0.3mm},
		font = {\fontsize{7pt}{12}\selectfont},
		linesplit/.style = {circle, draw = black, fill = black, inner sep = 0pt, minimum height = 2},
		linethick/.style = {-, line width = 0.60mm},
		linethin/.style={-,line width = 0.30mm},
	]

	\node[urec, minimum height=40.00, minimum width=40.0, text width=2cm, align=center, draw = color_semantic, rounded corners=5] (App1) at (-14,0) {Semantic RV\\ $\relvec\sim \prob(\relvec)$};
	\node[urec, minimum height=40.00, minimum width=40.0, text width=2cm, align=center, draw = color_semantic, rounded corners=5] (SChan) at (-11.25,0) {Semantic Channel\\ $\prob(\semvec|\relvec)$};
	\node[urec, minimum height=30.00, minimum width=60.0, text width=2.3cm, align=center, draw = color_communications_design, rounded corners=5] (Enc) at (-8,0) {Semantic Encoder\\ $\prob_{\txpars}(\xvec|\semvec)$};
	\node[urec, minimum height=30.00, minimum width=60.0, text width=2.3cm, align=center, draw = color_channel, rounded corners=5] (Chan) at (-8,-2) {Communication Channel $\prob(\yvec|\xvec)$};
	\node[urec, minimum height=30.00, minimum width=60.0, text width=2.3cm, align=center, draw = color_communications_design] (Dec) at (-8,-4.5) {Semantic Decoder\\ $\aprob_{\rxpars}(\relvec|\yvec)$};
	\node[urec, minimum height=30.00, minimum width=60.0, text width=2.3cm, align=center, draw = color_communications_design, rounded corners=5] (DM) at (-8,-6.5) {Decision-Making $\prob(\hrelvectech|\yvec)$};
	\node[urec, minimum height=40.00, minimum width=40.0, text width=2cm, align=center, draw = color_semantic] (sinfonydec) at (-11.25,-6.5) {Semantic Comm. Decision $\hrelvectech$};

	\node[urec, minimum height=40.00, minimum width=40.0, text width=2cm, align=center, draw = color_human_machine_interface, rounded corners=5] (HMI) at (-11.25,-3.25) {Semantics Presentation $\prob(\hdecvec|\yvec)$}; %
	\node[urec, minimum height=40.00, minimum width=50.0, text width=2.3cm, align=center, draw = color_human_decision_making, rounded corners=5] (HumDec) at (-14,-3.25) {Human Decision-Making $\prob(\hrelvec|\hdecvec)$};
	\node[urec, minimum height=40.00, minimum width=40.0, text width=2cm, align=center, draw = color_semantic] (HumDec2) at (-14,-6.5) {Human Decision\\ $\hrelvec$};

	\node[urec, inner xsep=4.5mm, inner ysep=4mm, text depth = 3 cm, fill=none, dashed, dash pattern={on 7pt off 3pt}, rounded corners=10, draw = color_communications_design, fit = (Enc) (DM), yshift=0.2cm, xshift=-0.3cm] (Enc2) {};
	\node[xshift = 0.0cm, yshift = -0.25cm, text width=3cm, align=center] at (Enc2.north) {\color{color_communications_design}{Communications Design}};
	\node[urec, inner ysep=3mm, inner xsep=1.5mm, text depth = 3 cm, fill=none, dashed, dash pattern={on 7pt off 3pt}, rounded corners=10, draw = color_semantic, fit = (App1) (SChan), yshift=0.15cm] (Semsource) {};
	\node[xshift = 0.0cm, yshift = -0.25cm, text width=3cm, align=center] at (Semsource.north) {\color{color_semantic}{Semantic Source $\prob(\semvec,\relvec)$}};

	\draw[arrowthick] (App1) -> (SChan) node[midway, above, rotate=0, align=center, anchor=south, yshift = 0.2cm] {$\relvec$};
	\draw[arrowthick] (SChan) -> (Enc) node[midway, above, rotate=0, align=center, anchor=south, yshift = 0.2cm] {$\semvec$};
	\draw[arrowthick] (Enc) -> (Chan) node[midway, above, rotate=0, align=center, anchor=west, yshift = 0.0cm, xshift=0.2cm] {$\xvec$};
	\node[linesplit, minimum height=5] (hnode) at (-8,-3.25) {};
	\draw[arrowthick] (hnode) -> (HMI);
	\draw[linethick] (Chan) -- (hnode)  node[pos=0.75, above, rotate=0, align=center, anchor=west, yshift = 0.0cm, xshift=0.2cm] {$\yvec$};
	\draw[arrowthick] (hnode) -> (Dec);
	\draw[arrowthick, dashedbox] (Dec.west) -- ++(-0.5,0) |- ([yshift=-0.4cm]HMI.east);
	\draw[arrowthick] (Dec) -> (DM) node[pos=0.70, above, rotate=0, anchor=south, yshift = 0.2cm] {};
	\draw[arrowthick] (HMI) -> (HumDec) node[pos=0.5, above, rotate=0, anchor=south, yshift = 0.2cm] {$\hdecvec$};
	\draw[arrowthick] (HumDec) -> (HumDec2) node[pos=0.5, right, rotate=0, anchor=west, yshift = 0.0cm, xshift=0.2cm] {$\hrelvec$};
	\draw[arrowthick] (DM) -> (sinfonydec) node[pos=0.5, right, rotate=0, anchor=west, yshift = 0.2cm, xshift=0.0cm] {$\hrelvectech$};

	\draw[arrowthick,<->,draw=none] (-15,1.4) -- (-7,1.4) node[midway, above, align=center, inner ysep=0.1cm, font = {\fontsize{9pt}{12}\selectfont},fill=none] {\textbf{E2E Sensing-Decision Framework:} $\prob\left(\relvec,\semvec,\xvec,\yvec,\hdecvec,\hrelvec\right)$};

\end{tikzpicture}%
    \caption{Block diagram of the end-to-end sensing-decision framework,~i.e., the probabilistic system model including human decision-making.}%
    \label{fig1:joint_sysmodel_human}
\end{figure}

To elaborate on our idea, we now describe our proposed end-to-end sensing-decision framework, which consists of multiple steps, exemplarily sketched in \reffig{fig0:e2e_tool_example} and modeled as shown in \reffig{fig1:joint_sysmodel_human}. It is based on the semantic communication model of~\cite{beck_semantic_2023}, including the complete communication Markov chain with the HDM model.

\subsection{Semantic Source}

The human performs tasks such as ensuring that machines in production run smoothly, which requires judging whether a tool is damaged or still operational. We will refer to this tool categorization task as our guiding example, whose flow is sketched in \reffig{fig0:e2e_tool_example}. The task defines the model of the world,~i.e., the semantics, and is described by a semantic multivariate Random Variable (\rv{}) $\relvec\in\relset^{\Nrelvec\times 1}$ from the domain $\relset$ of dimension $\Nrelvec$, distributed according to a probability density or mass function (pdf/pmf) $\prob(\relvec)$~\cite{beck_semantic_2023}. To simplify the discussion, we assume that it is discrete and memoryless.\footnote{For the rest of the article, note that the domain of all \rv{}s $\dom$ can be either discrete or continuous. Also note that the definition of entropy is different for discrete and continuous \rv{}s. For example, the differential entropy of continuous \rv{}s can be negative, while the entropy of discrete \rv{}s is always positive~\cite{simeone2018brief}. Thus, without loss of generality, we will assume that all \rv{}s are either discrete or continuous. In this paper, we avoid notational clutter by using the expectation operator: By replacing the summation over discrete \rv{}s with an integral, the equations are valid for continuous \rv{}s, and vice versa~\cite{beck_semantic_2023}.}
The semantic source $\prob(\semvec,\relvec)$ links the semantics expressed by $\relvec$ with the sensed, observed signal \rv{} $\semvec\in\semset^{\Nsemvec\times 1}$ that enters the communication system. This sensed data can be,~e.g., images of a tool taken from different perspectives, as shown in \reffig{fig0:e2e_tool_example}. The semantic link can be modeled in the Markov chain by a semantic channel, a conditional distribution $\prob(\semvec|\relvec)$, as shown in \reffig{fig1:joint_sysmodel_human}.

\subsection{Semantic Communication}%
\label{sec:semantic_communication}

The semantic communication system encodes the sensed signal $\semvec$ with the encoder $\prob_{\txpars}(\xvec|\semvec)$, parametrized by $\txpars\in\realnum^{\Ntxpars\times 1}$, to the transmit signal $\xvec\in\xset^{\ntx\times 1}$ (see \reffig{fig1:joint_sysmodel_human}) for efficient and reliable semantic transmission over the physical communication channel $\prob(\yvec|\xvec)$, where $\yvec\in\yset^{\Nrx\times 1}$ is the received signal vector, so that the semantic \rv{} $\relvec$ is best preserved~\cite{beck_semantic_2023}. At the receiver side, the decoder $\aprob_{\rxpars}(\relvec|\yvec)$ with parameters $\rxpars\in\realnum^{\Nrxpars\times 1}$ recovers the semantics $\relvec$ for the receiver.

In~\cite{beck_semantic_2023}, the authors identified the Information Maximization (InfoMax) problem as an appropriate design criterion for semantic communication, since it maximizes the amount of mutual information $\mi[\txpars]{\relvec;\yvec}$ of the semantic \rv{} $\relvec$ contained in the received signal $\yvec$:
\begin{align}
       & \argmax[\prob_{\txpars}(\xvec|\semvec)]\mi[\txpars]{\relvec;\yvec} \label{eq:infomax}                                                                                                                          \\
    =~ & \argmax[\txpars]\eval[\relvec,\yvec\sim\prob_{\txpars}(\relvec,\yvec)]{\ln \frac{\prob_{\txpars}(\relvec,\yvec)}{\prob(\relvec)\prob_{\txpars}(\yvec)}}                                                        \\
    =~ & \argmax[\txpars]\HE[\relvec] - \HEN[\txpars]{\relvec|\yvec}                                                                                                                                                    \\
    =~ & \argmax[\txpars]\eval[\relvec,\yvec\sim\prob_{\txpars}(\relvec,\yvec)]{\ln \prob_{\txpars}(\relvec|\yvec)} \label{eq:infomax2}                                                                                 \\
    =~ & \argmax[\txpars]\eval[\relvec,\semvec,\xvec,\yvec\sim\prob(\yvec|\xvec)\cdot\prob_{\txpars}(\xvec|\semvec)\cdot \prob(\semvec,\relvec)]{\ln \prob_{\txpars}(\relvec|\yvec)} \eqpoint \label{eq:markov_infomax}
\end{align}
There, $\evalnb[\xvec\sim \prob(\xvec)]{\funcrtrick(\xvec)}$ denotes the expected value of $\funcrtrick(\xvec)$ with respect to both discrete and continuous \rv{} $\xvec$, $\HE[\relvec]=\evalnb[\relvec\sim\prob(\relvec)]{-\ln\prob(\relvec)}$ the entropy of $\relvec$, and $\HE[\relvec|\yvec]$ the conditional entropy.

If the computation of the posterior $\prob_{\txpars}(\relvec|\yvec)$ in~\eqref{eq:infomax2} is intractable, we can replace it by a variational distribution,~i.e., the decoder $\aprob_{\rxpars}(\relvec|\yvec)$, to define a Mutual Information Lower Bound (MILBO)~\cite{vincent_stacked_2010,simeone2018very,beck_semantic_2023}:
\begin{align}
    \mi[\txpars]{\relvec;\yvec} & \geq \HE[\relvec] + \eval[\relvec,\yvec\sim\prob_{\txpars}(\relvec,\yvec)]{\ln \aprob_{\rxpars}(\relvec|\yvec)} \label{eq:milbo}              \\
                                & =\HE[\relvec] + \eval[\yvec\sim\prob_{\txpars}(\yvec)]{\eval[\relvec\sim\prob_{\txpars}(\relvec|\yvec)]{\ln \aprob_{\rxpars}(\relvec|\yvec)}} \\
                                & = \HE[\relvec] -\ceparams  \eqpoint \label{eq:ce_loss_opt}
\end{align}

Noting that only the negative amortized cross-entropy $\ceparams$ in~\eqref{eq:ce_loss_opt} depends on both $\txpars$ and $\rxpars$, and fixing the transmit dimension to $\ntx$, we can optimize encoder and decoder parameters~\cite{beck_semantic_2023}:
\begin{align}
    \lcb\txpars^*,\rxpars^*\rcb & =\argmin[\txpars,\rxpars] \ceparams \eqpoint \label{eq:ce_opt}
\end{align}
Note that the form of $\prob_{\txpars}(\yvec|\semvec)$ must be constrained to avoid learning a trivial identity mapping $\yvec=\semvec$. In fact, we constrain the optimization and information rate by our communication channel $\prob(\yvec|\xvec)$ and number of channel uses $\ntx$, which we assume to be given. This introduces an Information Bottleneck (IB). Alternatively, we can explicitly constrain the information rate $\mi[\txpars]{\semvec;\yvec}$ in an IB problem~\cite{beck_semantic_2023, beck_model-free_2024}. To solve~\eqref{eq:ce_opt}, we use the empirical cross-entropy and ML techniques such as DNNs, Stochastic Gradient Descent (SGD), and the reparametrization trick to obtain our ML-based design Semantic INFOrmation TraNsmission and RecoverY (\sinfoni{})~\cite{beck_semantic_2023, beck_model-free_2024}.

Through the discriminative modeling assumption $\prob_{\txpars}(\xvec|\semvec)$, sample-based optimization, and the use of a variational decoder, we avoid explicit modeling of the high-dimensional semantic source distribution $\prob(\semvec,\relvec)$ and computation of high-dimensional integrals for mutual information computation --- crucial benefits. To handle high-dimensional data $\semvec$, we need to use more expressive models, such as DNNs, to accurately approximate the posterior with the variational decoder $\aprob_{\rxpars}(\relvec|\yvec)$. In semantic communication, we can also exploit the encoder $\prob_{\txpars}(\xvec|\semvec)$ to reduce the dimensionality facilitating effective decoder processing and saving bandwidth, enabled by the assumption $\HE[\semvec]\ge\HE[\relvec]$~\cite{beck_semantic_2023}. Care must be taken, however, to avoid losing relevant information about $\relvec$ in $\yvec$,~i.e., to maintain high $\mi[\txpars]{\relvec;\yvec}$.

We note that the semantic communication system is able to make a decision by itself after optimization/training based on the decoder $\aprob_{\rxpars}(\relvec|\yvec)$. For the discrete \rv{}s, the most likely option,~i.e., the Maximum A-Posteriori (MAP) estimate, is optimal:
\begin{align}
    \hrelvectech=\argmax[\relvec] \aprob_{\rxpars}(\relvec|\yvec) \eqpoint \label{eq:sinfony_decision}
\end{align}
This decision process in operation mode is modeled as $\prob(\hrelvectech|\yvec)$ (see \reffig{fig1:joint_sysmodel_human}).

\subsection{Semantics Presentation}%
\label{sec:semantics_presentation}

Finally, semantic communication presents the received signal $\yvec$ or the extracted probabilistic semantic decoder estimate $\aprob_{\rxpars}(\relvec|\yvec)$ --- in the best case containing maximum amount of information about the semantic \rv{} $\relvec$ according to~\eqref{eq:infomax2} or~\eqref{eq:milbo} --- to the HDM model. We describe this process by $\prob(\hdecvec|\yvec)$ with a presentation \rv{} $\hdecvec\in\realnum^{\Nfinalfeat\times 1}$. In practice, the presentation $\hdecvec$ must be tailored to a human, requiring a Human-Machine Interface (HMI) that is typically designed handcrafted, such as visualization (see \reffig{fig0:e2e_tool_example}). In this work, we abstract the HMI as in a technical system --- where the components are connected by a deterministic function $\hdecvec=\func{\yvec}$.

\subsection{Human Decision-Making Model}%
\label{sec:hdm_model}

Based on the HMI or semantics presentation $\hdecvec$, the human decision-maker then makes a decision to complete the overall task. In this work, we will model the Human Decision-Making (HDM) process probabilistically by $\prob(\hrelvec|\hdecvec)$ to make a first step towards integrating and evaluating the human with the technical system,~i.e., semantic communication and HDM\@. Finally, by decision, we obtain the estimated semantics $\hrelvec\in\relset^{\Nrelvec\times 1}$, which can be different from the true semantic \rv{} $\relvec$ (see \reffig{fig0:e2e_tool_example}). In our guiding example, this could mean that the HDM process decides that the tool is damaged even though it is still usable, and vice versa.

Reflecting the variance in decision tasks, the literature on HDM includes a variety of theoretical models and approaches to capture decision processes~\cite{koehler2008blackwell,newell2022straight}. The most appropriate model often varies depending on the type of decision task and context. In this example, we focus on categorization tasks where the decision-maker must decide based on their experience whether an object belongs to one of $\Nclass$ categories, such as whether a tool can still be used or whether the concentration of a toxic gas is above a certain threshold.

\subsubsection{Generalized Context Model}

While a large number of increasingly complex models of human categorization have been proposed~\cite{anderson_adaptive_1991, kurtz_chapter_2015, schlegelmilch_cognitive_2022}, the core assumptions of the Generalized Context Model (GCM)~\cite{nosofsky_choice_1984,wills_generalized_2011} are commonly adapted by many successors~\cite{nosofsky_exemplar-based_2015} and have been successfully used to describe categorizations of complex real world stimuli~\cite{bhatia_naturalistic_2018, meagher_testing_2023}.

Despite the relative simplicity of the GCM, the well-studied model and its variants can easily account for HDM under different contexts,~e.g., under time pressure~\cite{lamberts1995categorization, seitz2023testing}, capture human judgment under cognitive load~\cite{hoffmann2013deliberation}, and explain common HDM biases,~e.g., base-rate bias~\cite{shanks1991connectionist}. At the same time, GCM has been applied to different aspects of human cognition,~e.g., artificial grammar~\cite{pothos2000role}, leadership competence judgment~\cite{sewell2022exemplifying}, and mental multiplication~\cite{griffiths2002multidimensional}. Several extensions of GCM have been created to explain even broader aspects of HDM such as reaction time in decision-making~\cite{nosofsky1997exemplar} or learning~\cite{lee2002extending}. Since the GCM is a powerful approximation to human categorization decisions, we choose it to simulate the decision-making process.

An important assumption of the GCM is that categorization decisions are made on the basis of exemplar memory,~i.e., previously seen realizations that are retrieved from memory. Accordingly, in the tool example, the model assumes that the decision-maker first experiences $\Nsamp$ tool realizations and whether those tools need to be replaced. These tools are then remembered as the $\idxi$-th ``exemplar'',~i.e., realization $\hdecvecreal_{\idxi}$, with the corresponding label $\relvecreal_{\idxi}$, so we have an exemplar dataset or HDM knowledge base $\datasetexemplar={\{\hdecvecreal_{\idxi},\relvecreal_{\idxi}\}}_{\idxi=1}^{\Nsamp}$. Since the semantic \rv{} $\relvec$ is a categorical \rv{}, we can describe it by a one-hot vector $\relvec=\ohfunc(\relvarvalue)$ where all elements are zero except for the element $\relvarvalue\in\{1,\ldots,\Nclass\}$ that represents the tool state from a total number of $\Nclass$ states. For example, for binary states, we have $\relvarvalue\in\{1,2\}$ with $\Nclass=2$.

When the decision-maker encounters a new tool presentation $\hdecvec$, the probability $\aprob_{\paramsgcm}(\relvec=\relvecreal|\hdecvec,\datasetexemplar)=\aprob(\relvec=\relvecreal|\hdecvec,\datasetexemplar,\paramsgcm)$ of making the decision $\relvecreal=\ohfunc(\relvarvaluereal)$ given this representation $\hdecvec$ is the result of the comparison between $\hdecvec$ and all seen realizations $\hdecvecreal_{\idxi}$ from $\datasetexemplar$:
\begin{align}
    \aprob_{\paramsgcm}(\relvec=\relvecreal|\hdecvec,\datasetexemplar) = \frac{\sum \limits_{\idxi=1}^{\Nsamp} \similarity{\hdecvecreal_{\idxi}}{\hdecvec|\paramsgcm} \cdot \iversonbracket[\relvecreal_{\idxi}=\relvecreal]}{\sum \limits_{\idxi=1}^{\Nsamp} \similarity{\hdecvecreal_{\idxi}}{\hdecvec|\paramsgcm}} \label{eq:gcm_model}
\end{align}
with GCM parameters $\paramsgcm$ and $\iversonbracket[\relvecreal_{\idxi}=\relvecreal]$ being the Iverson bracket, which is equal to $1$ if $\relvecreal_{\idxi}=\relvecreal$ and $0$ otherwise. This means the approximating posterior $\aprob_{\paramsgcm}(\relvec|\hdecvec,\datasetexemplar)$ is determined by the sum of similarities between $\hdecvec$ and all the seen realizations $\relvecreal_{\idxi}$ that belong to the decision $\relvecreal$, and normalized by the similarity to all the seen realizations regardless of the decision. We note that the model~\eqref{eq:gcm_model} assumes that the decision-maker has perfect memory of its knowledge base $\datasetexemplar$.

The similarity $\similarity{\hdecvecreal_{1}}{\hdecvecreal_{2}|\paramsgcm}$ between two presentations decreases exponentially as the Euclidean distance between two presentations increases and depends on the learnable GCM parameters $\paramsgcm=\{\simconst,\simweightvec\}$:
\begin{align}
    \similarity{\hdecvecreal_{1}}{\hdecvecreal_{2}|\paramsgcm} & = e^{-\simconst \cdot \distfunc{\hdecvecreal_{1},\hdecvecreal_{2}|\simweightvec}} \nonumber                                                                                                  \\
                                                               & = e^{-\simconst \cdot {\left(\abs[\hdecvecreal_{1}-\hdecvecreal_{2}]^{\trapo}\cdot\diag{\simweightvec}\cdot\abs[\hdecvecreal_{1}-\hdecvecreal_{2}]\right)}^{1/2}}  \label{eq:gcm_similarity}
\end{align}
with $\diag{\simweightvec}$ creating a diagonal matrix with elements of $\simweightvec\in\realnum^{\Nfinalfeat\times 1}$ on its diagonal and distance measure $\distfunc{\hdecvecreal_{1},\hdecvecreal_{2}|\simweightvec}$.

A unique assumption of the GCM is that each element or feature $\hdecvarreal_{\idxn}$ of a seen realization $\hdecvecreal_{\idxi}$ is weighted by attention weights $\simweight_{\idxn}$ and hence does not contribute equally to the perceived similarity. To achieve the attention functionality, the weights are constrained by $\simweight_{\idxn}\ge 0$ and normalized by $\sum_{\idxn=1}^{\Nfinalfeat} \simweight_{\idxn}=1$. The parameter $\simconst$ has two interpretations: First, the similarity gradient $\simconst$ describes the sharpness of the decline in similarity, with higher $\simconst$ resulting in a sharper decline of similarity when the distance increases. Second, the parameter $\simconst$ describes the consistency in making decisions and reflects the probabilistic nature of the HDM process, akin to the temperature parameter in the Boltzmann distribution~\cite{luce_possible_1959}. Accordingly, a lower parameter $\simconst$ results in a new tool being more confidently categorized to the category with higher similarity.

\subsubsection{HDM-based Probabilistic Decision-Making}

After training of the GCM (see \refsec{sec:hdm_optimization}), the strategy of the HDM model is equal to the random process~\cite{friedman_monty_1998, sanborn_noise_2024}:
\begin{align}
    \hrelvec\sim\prob(\hrelvec|\hdecvec)=\aprob_{\paramsgcm}(\relvec=\hrelvec|\hdecvec,\datasetexemplar) \eqpoint \label{eq:gcm_decision}
\end{align}
When faced with options of varying probabilities, people tend to distribute their choices according to the probability distribution rather than always selecting the most likely option resulting in a suboptimal policy~\cite{friedman_monty_1998, sanborn_noise_2024}. Thus, the accuracy of human decisions can be worse compared to that of the optimal deterministic policy~\eqref{eq:sinfony_decision} of the technical (semantic communication) system,~e.g., \sinfoni{}.

\subsection{End-to-End Sensing-Decision Model}

With all the aforementioned subcomponent models, we are able to create a generative model of the end-to-end sensing-decision process answering the first research question \emph{a) Joint Modeling}. We note that we can distinguish between four different models corresponding to four system stages:
\begin{enumerate}
    \item Design of semantic encoder $\prob_{\txpars}(\xvec|\semvec)$ and decoder $\aprob_{\rxpars}(\relvec|\yvec)$ based on the forward communications model
          \begin{align}
              \prob(\relvec,\semvec,\xvec,\yvec)=\prob(\relvec,\semvec)\cdot\prob_{\txpars}(\xvec|\semvec)\cdot\prob(\yvec|\xvec) \eqpoint \label{eq:_sinfony_design_model}
          \end{align}
    \item Semantic communication is executed in operation mode to make decisions via~\eqref{eq:sinfony_decision}. Then, the model is
          \begin{align}
              \prob(\relvec,\semvec,\xvec,\yvec,\hrelvectech)=\prob(\relvec,\semvec,\xvec,\yvec)\cdot\prob(\hrelvectech|\yvec) \eqpoint \label{eq:sinfony_e2e_model}
          \end{align}
    \item The HDM model $\aprob_{\paramsgcm}(\relvec|\hdecvec,\datasetexemplar)$ is trained based on seen presentation and label realizations from semantic communication in operation mode. The underlying model is
          \begin{align}
              \prob(\relvec,\semvec,\xvec,\yvec,\hdecvec)=\prob(\relvec,\semvec,\xvec,\yvec)\cdot\prob(\hdecvec|\yvec) \eqpoint \label{eq:hdm_training_model}
          \end{align}
    \item Semantic communication presents information to the HDM model that finally makes a decision. The overall end-to-end sensing-decision model of \reffig{fig1:joint_sysmodel_human} in operation mode after all training phases is:
          \begin{align}
              \prob(\relvec,\semvec,\xvec,\yvec,\hdecvec,\hrelvec)=\prob(\relvec,\semvec,\xvec,\yvec,\hdecvec)\cdot \prob(\hrelvec|\hdecvec) \eqpoint \label{eq:e2e_model}
          \end{align}
\end{enumerate}
We emphasize the broad applicability of the proposed end-to-end sensing-decision framework, as it relies on generic Markov chain and probabilistic modeling assumptions, including the HDM decision, and can naturally be extended to incorporate,~e.g., sensing, timing and power aspects such as in~\cite{wen_task-oriented_2024}. Since any semantically meaningful information can be communicated to the HDM model regardless of implementation details, the framework and its conclusions are widely transferable.

\subsection{Information-theoretic Overall View on Design in the End-to-End Sensing-Decision Framework}%
\label{sec:infomax_extension}

We can exploit our end-to-end sensing-decision framework~\eqref{eq:_sinfony_design_model}-\eqref{eq:e2e_model}, to extend the information-theoretic view of semantic communication to both the semantics presentation and the HDM model to gain new insights.

\subsubsection{HDM Model -- Training}%
\label{sec:hdm_optimization}

In this work, the GCM is optimized given a presentation $\hdecvec$ based on a fixed, optimized semantic communication system of model~\eqref{eq:hdm_training_model}. In general, the GCM parameters $\paramsgcm=\{\simconst,\simweightvec\}$ are optimized using the maximum likelihood criterion~\cite{wills_generalized_2011}. We note that maximization of the log-likelihood function is equal to amortized minimization of the empirical cross-entropy on the training set~\cite{simeone2018brief}. Transferring the InfoMax view from the semantic communication system in~\eqref{eq:ce_loss_opt}, this means we optimize a lower bound on the mutual information $\mi[\txpars]{\relvec;\hdecvec}$, but now between $\relvec$ and $\hdecvec$ with respect to \paramsgcm:
\begin{align}
    \mi[\txpars]{\relvec;\hdecvec} & \geq \HE[\relvec] + \eval[\relvec,\hdecvec\sim\prob_{\txpars,\rxpars}(\relvec,\hdecvec)]{\ln \aprob_{\paramsgcm}(\relvec|\hdecvec,\datasetexemplar)} \\
                                   & = \HE[\relvec]-\Lossf_{\paramsgcm}^{\mathtxt{CE}} \eqpoint \label{eq:ce_gcm}
\end{align}
From an information-theoretic view, we conclude that the choice of the GCM optimization criterion
\begin{align}
    \paramsgcm^* & =\argmin[\paramsgcm] \Lossf_{\paramsgcm}^{\mathtxt{CE}} \label{eq:ce_gcm_opt}
\end{align}
is well-motivated. Furthermore, comparing~\eqref{eq:milbo} and~\eqref{eq:ce_gcm}, we conclude that \emph{the HDM model can be interpreted as a decoder}. If the GCM is the decoder, then we can show --- using a Markov chain decomposition as in~\eqref{eq:markov_infomax} --- that we optimize it \emph{given a new overall encoder} $\prob_{\txpars,\rxpars}(\hdecvec|\semvec)$ including semantic communication and semantics presentation (see \reffig{fig1:joint_sysmodel_human}). To solve~\eqref{eq:ce_gcm_opt}, computer search methods such as the differential annealing algorithm are typically used~\cite{wills_generalized_2011,Zhang2012DifferentialAnnealing}. In this work, for computational efficiency and joint integration with \sinfoni{}, we implemented the GCM in the automatic differentiation framework TensorFlow and exploit SGD for optimization of its parameters until training convergence. The source code is available in~\cite{Beck_Semantic_INFOrmation_traNsmission_2024}.

We note that training rarely achieves full convergence in real-world environments. For example, humans seldom attain full mastery of tasks due to limited access to training samples and repetitions, as well as variable conditions such as motivation and distractions~\cite{meagher_testing_2023}. Consequently, simulations typically provide an upper bound on HDM performance.
Moreover, training suboptimality is reflected in a property of~\eqref{eq:ce_gcm_opt}: The objective function $\Lossf_{\paramsgcm}^{\mathtxt{CE}}$ is non-convex in the GCM parameters $\paramsgcm$. The proof is provided in the Appendix~\ref{sec:non_convexity}. Consequently, iterative optimization algorithms are not guaranteed to reach the global minimum; we therefore adopt strategies from DNN training to converge to a good local minimum. To regularize the search space, we choose the starting point $\simconst=1$ and $\simweightvec=\ones/\Nfinalfeat$, which gives equal importance to all presentation variables from the start.

\subsubsection{Semantics Presentation -- Design Optimization}%
\label{sec:presentation_optimization}

Moreover, if we make the presentation process $\prob_{\paramspres}(\hdecvec|\yvec)$ with parameters $\paramspres$ tunable and add the resulting overall encoder $\prob_{\paramsoverall}(\hdecvec|\semvec)$ with parameters $\paramsoverall=\{\txpars,\rxpars,\paramspres\}$ to the optimization problem~\eqref{eq:ce_gcm_opt}, we arrive at the MILBO objective function:
\begin{align}
    \mi[\paramsoverall]{\relvec;\hdecvec} & \geq \HE[\relvec] \nonumber                                                                                                                     \\ &\quad + \eval[\relvec,\yvec,\hdecvec\sim\prob_{\txpars,\rxpars}(\relvec,\yvec)\cdot\prob_{\paramspres}(\hdecvec|\yvec)]{\ln \aprob_{\paramsgcm}(\relvec|\hdecvec,\datasetexemplar)} \nonumber \\
                                          & =                                           \HE[\relvec] -\Lossf_{\paramsoverall,\paramsgcm}^{\mathtxt{CE}} \eqpoint \label{eq:ce_presentation}
\end{align}
Now, minimizing the amortized cross-entropy,~i.e.,
\begin{align}
    \{\paramsoverall^{*},\paramsgcm^{*}\} = \argmin[\paramsoverall,\paramsgcm] \Lossf_{\paramsoverall,\paramsgcm}^{\mathtxt{CE}} \eqcomma \label{eq:ce_joint_opt}
\end{align}
allows for joint end-to-end optimization of all components with respect to all framework parameters $\txpars,\rxpars,\paramspres,\paramsgcm$, leading to a unified design. We note that we can solve~\eqref{eq:ce_joint_opt} given a fixed semantic communication system to only tune the presentation $\prob_{\paramspres}(\hdecvec|\yvec)$, which we first focus on.

Decomposing the amortized cross-entropy $\Lossf_{\paramsoverall,\paramsgcm}^{\mathtxt{CE}}$ as in~\cite{beck_semantic_2023} into
\begin{align}
    \Lossf_{\paramsoverall,\paramsgcm}^{\mathtxt{CE}} & = \HE[\relvec] - \mi[\paramsoverall]{\relvec;\hdecvec}  \label{eq:theoretical_design_criteria}                                                                                \\
                                                      & \quad + \eval[\hdecvec\sim\prob_{\paramsoverall}(\hdecvec)]{\dkl{\prob_{\paramsoverall}(\relvec|\hdecvec)}{\aprob_{\paramsgcm}(\relvec|\hdecvec,\datasetexemplar)}} \nonumber
\end{align}
reveals two possibly conflicting design criteria:
\begin{enumerate}
    \item The presentation encoder $\prob_{\paramspres}(\hdecvec|\yvec)$ should maximize the mutual information $\mi[\paramsoverall]{\relvec;\hdecvec}$ that depends solely on it through the true posterior $\prob_{\paramsoverall}(\relvec|\hdecvec)$ (see~\eqref{eq:infomax2}).
    \item Both true overall posterior $\prob_{\paramsoverall}(\relvec|\hdecvec)$ --- and hence the presentation encoder $\prob_{\paramspres}(\hdecvec|\yvec)$ --- and the HDM model $\aprob_{\paramsgcm}(\relvec|\hdecvec,\datasetexemplar)$ are matched by minimizing the Kullback-Leibler (KL) divergence.
\end{enumerate}
In a technical semantic communication system from \refsec{sec:semantic_communication}, we can avoid the design conflict by using a model $\aprob_{\rxpars}(\relvec|\yvec)$ expressive enough to approximate $\prob_{\txpars}(\relvec|\yvec)$ arbitrarily well, such that the focus lies on the InfoMax term. However, in case of the end-to-end sensing-decision training model~\eqref{eq:hdm_training_model}, if the HDM model (or human) constrains the form (or capacity) of $\aprob_{\paramsgcm}(\relvec|\hdecvec,\datasetexemplar)$,~i.e., the solution space, to some degree, the two optimization terms in~\eqref{eq:theoretical_design_criteria} are traded off: Then, the true posterior $\prob_{\paramsoverall}(\relvec|\hdecvec)$ has to be fit to $\aprob_{\paramsgcm}(\relvec|\hdecvec,\datasetexemplar)$ and we do not maximize $\mi[\paramsoverall]{\relvec;\hdecvec}$ alone which could lead to a loss in mutual information.

\emph{Example 1:} These abstract information-theoretic insights explain well what we observe in practice with handcrafted presentations. In reality, it is difficult to understand and subsequently visualize the received raw communications signal $\yvec$ for a human without any preprocessing:
\begin{itemize}
    \item We have to match the presentation encoder $\prob_{\paramspres}(\hdecvec|\yvec)$ to the cognitive process $\aprob_{\paramsgcm}(\relvec|\hdecvec,\datasetexemplar)$ according to the KL divergence term in~\eqref{eq:theoretical_design_criteria}. Fortunately, the semantic decoder $\aprob_{\rxpars}(\relvec|\yvec)$ obtained by maximizing the MILBO extracts the semantic information of $\relvec$ from $\yvec$ and allows for meaningful presentation to and interpretation by the HDM model $\aprob_{\paramsgcm}(\relvec|\hdecvec,\datasetexemplar)$ (or human).
    \item However, we may lose relevant information about $\relvec$ fitting the presentation to the cognitive process including the semantic decoder preprocessing in the Markov chain $\relvec\rightarrow\yvec\rightarrow \aprob_{\rxpars}(\relvec|\yvec)\rightarrow \hdecvec$ according to the data processing inequality
          \begin{align}
              \mi{\relvec;\yvec} \ge\mi{\relvec;\aprob_{\rxpars}\left(\relvec|\yvec\right)}\ge \mi{\relvec;\hdecvec} \eqpoint \label{eq:data_processing_inequality}
          \end{align}
\end{itemize}
\emph{Example 2:} Another example of how the HDM process influences presentation design is that research on HDM models focuses on the interplay between relevant features $\hdecvec$ and requires a certain level of feature extraction $\prob_{\paramspres}(\hdecvec|\semvec)$ from raw images $\semvec$ (or $\yvec$) for HDM model processing. For an overview of this research, we refer the reader to~\cite{serre2016models}. These HDM models were not built to process raw images $\semvec$ directly,~i.e., $\aprob_{\paramsgcm}(\relvec|\hdecvec=\semvec,\datasetexemplar)$, which would lead to unrealistically poor performance despite maximum relevant information in $\semvec$ about $\relvec$. Thus, in the numerical results of \refsec{sec:end2end_simulation}, we cannot compare to a setup where the raw data of the images $\semvec$ are digitally communicated and then directly processed by the HDM model.

Based on our end-to-end sensing-decision framework, we conclude that it highly depends on the processing capabilities of the HDM model if it can extract more or less information about $\relvec$ from $\yvec$ than the semantic decoder. Moreover, we conclude that balancing of two possibly conflicting criteria is key for presentation design:
\begin{enumerate}
    \item \textbf{Relevant information preservation:} On the one hand, careful design of $\hdecvec$ is required to not lose any relevant information about $\relvec$ for the final decision. For example, the higher the dimension $\Nfinalfeat$ of the presentation \rv{} $\hdecvec$, the more detailed the presentation to the HDM model is, and the more information it contains.
    \item \textbf{Presentation alignment to the HDM model:} On the other hand, the presentation has to be in a form that can be understood by the HDM model, effectively restricting the set of possible presentations $\hdecvec$. For example, compressing the relevant information about $\relvec$ into $\hdecvec$ may be required to ease cognitive processing, thereby enhancing both the effectiveness (decision accuracy) and efficiency (response speed).
\end{enumerate}
To investigate how to balance these two design rules, we compare two types of handcrafted presentations in our numerical example of \refsec{sec:presentation_design}, which we extend into end-to-end optimized presentations in \refsec{sec:simulation_scenarios}. For an outlook on the inclusion of HMIs in practice, please refer to \refsec{sec:outlook}.

\section{Simulative Investigation}%
\label{sec:simulative_investigation}

In this section, we evaluate first numerical results of our joint framework using classification examples. The inclusion of diverse datasets for semantic source modeling, such as the standard MNIST and CIFAR10 datasets and audio stimuli in addition to our guiding tool example, enables the generalization of conclusions beyond the specific case of tool wear.

\subsection{Performance Measures}

We measure the performance of decision-making for both semantic communication and the HDM model by the categorical accuracy
\begin{align}
    \acc = \eval[\relvec\sim\prob(\relvec)]{\prob\left(\hrelvectech=\relvec|\relvec\right)} & =\summ{\relvec\in\relset^{\Nrelvec\times 1}} \prob\left(\hrelvectech=\relvec,\relvec\right)                                      \\
                                                                                            & \approx \frac{1}{\Nsamp}\summ[\Nsamp]{\idxi=1} \iversonbracket[\hrelvectechreal_{\idxi}=\relvecreal_{\idxi}] \label{eq:accuracy}
\end{align}
--- comparing predicted and true category realizations $\hrelvectechreal_{\idxi}$ and $\relvecreal_{\idxi}$ --- or the classification error rate $1-\acc$ common in communications. Since the HDM model decides probabilistically based on the input $\hdecvecreal_{\idxi}$, we can calculate the accuracy based on the end-to-end sensing-decision model~\eqref{eq:e2e_model} shown in \reffig{fig1:joint_sysmodel_human} by the sum of the probabilities of the GCM responding to the correct category $\relvecreal_{\idxi}$~\cite{nosofsky_choice_1984}:
\begin{align}
    \acc & = \eval[\relvec\sim\prob(\relvec)]{\prob\left(\hrelvec=\relvec|\relvec\right)}                                                                          \\
         & =       \eval[\relvec\sim\prob(\relvec)]{\eval[\hdecvec\sim\prob(\hdecvec|\relvec)]{\prob\left(\hrelvec=\relvec|\hdecvec\right)}}                       \\
         & =       \eval[\relvec,\hdecvec\sim\prob(\relvec,\hdecvec)]{\prob\left(\hrelvec=\relvec|\hdecvec\right)}                                                 \\
         & \approx \frac{1}{\Nsamp}\summ[\Nsamp]{\idxi=1} \prob(\hrelvec=\relvecreal_{\idxi}|\hdecvec=\hdecvecreal_{\idxi}) \label{eq:prob_matching}               \\
         & =       \frac{1}{\Nsamp}\summ[\Nsamp]{\idxi=1} \aprob_{\paramsgcm}(\relvec=\relvecreal_{\idxi}|\hdecvec=\hdecvecreal_{\idxi},\datasetexemplar) \eqpoint
\end{align}
This method of calculating accuracy is commonly used in psychology studies for HDM models~\cite{nosofsky_choice_1984}.

\subsection{Exemplary Semantic Source Datasets}

Tool wear and tool replacement decisions pose a common challenge in the metal cutting industry to reduce production costs~\cite{Prasad2001}, and represent an exemplary semantic source of this work along the ones summarized in \reftable{tab:datasets}.

\textbf{Tool Wear:} In this decision-making problem, the semantic \rv{} $\relvec$ is modeled as a binary variable with two states, where $\relvecreal = {[1,0]}^{\trapo}$ indicates a worn tool and $\relvecreal = {[0,1]}^{\trapo}$ indicates a usable tool. Although optical measurement techniques provide accurate assessments of tool wear, small and medium-sized companies often rely on machine operators to manually assess tool wear. To automate this process, a dataset was recorded where human experts were presented two different grayscale images of each tool~\cite{Papenberg_24}: One image $\semvec_{1}\in\pixelset^{218\times 380\times 1}$ with $\pixelset={\{0,1,\dots,255\}}$ taken from the side and another $\semvec_{2}\in\pixelset^{487\times 380\times 1}$ taken from the top (see \reffig{fig0:e2e_tool_example}). Based on these observations $\semvec=\{\semvec_{1}, \semvec_{2}\}$, the experts labeled the tools into the binary states $\relvec$.

\begin{table}[!t]
    \caption{Datasets -- Exemplary Semantic Sources}
    \label{tab:datasets}
    \setlength{\tabcolsep}{3pt}
    \centering
    \begin{tabular}{llrrr}
        \hline
        Dataset         & Type  & Classes & Size      & Dimension                         \\
        \hline
        Tool Wear       & Image & $2$     & $1,632$   & $(218,380), (487,380)$            \\
        MNIST           & Image & $10$    & $70,000$  & $(28,28,1)$                       \\
        CIFAR10         & Image & $10$    & $60,000$  & $(32,32,3)$                       \\
        Speech Commands & Audio & $12$    & $100,503$ & $16$ k $\rightarrow (124,129)$    \\
        UrbanSound8K    & Audio & $10$    & $8,732$   & $176.4$ k $\rightarrow (249,257)$ \\
        \hline
    \end{tabular}
\end{table}

This process resulted in a dataset $\tset = \lcb \left(\semvarreal_{\idxi}, \relvarreal_{\idxi} \right)\rcb_{\idxi=1}^{\Nsamp}$ modeling our semantic source $\prob(\semvec,\relvec)$ and consisting of $\Nsamp = 1,632$ data pairs, with an $85\%$ split between training and validation data.

\textbf{Classic Image Data:} We also revisit the MNIST and CIFAR10 examples from~\cite{beck_semantic_2023, Beck_Semantic_INFOrmation_traNsmission_2024} to extend our analysis to different tasks and larger datasets with $\Nsamp = 70,000$ and $\Nsamp = 60,000$, respectively.

For clarity and reproducibility, we select image classification as a representative example task. This choice is motivated by the fact that image classification is well understood, widely adopted in prior works, and allows for fair comparison under controlled conditions. In this way, the focus remains on the framework itself rather than on improving a particular downstream task.

\textbf{Audio Signals:} Moreover, we broaden the generality of our framework and its numerical results by testing on the modality of audio stimuli with the \emph{Speech Commands}~\cite{warden_speech_2018} and \emph{UrbanSound8K}~\cite{salamon_dataset_2014} datasets for keyword spotting and acoustic scene classification, see \reftable{tab:datasets}, respectively.

\textbf{Audio Preprocessing:} To preprocess the data samples as input for \sinfoni{}, we take the following steps: For UrbanSound8K, we downsample the signal from $44.1$ kHz to $16$ kHz and zero-pad to the maximum duration of $4$ s, reducing maximum sample size $176,400$ to uniform $64,000$-length samples. For $16$ kHz Speech Commands signals of $1$ s, we compute the spectrogram by the short-term Fourier transform with frame length of $255$ and frame steps of $128$ leading to a spectrogram size of $124\times 129$ and, for UrbanSound8K, $512$ and $256$ leading to $249\times257$, respectively. Lastly, we normalize the spectrograms by maximum and minimum value to $[0,1]$.

\subsection{Semantic Communication Analysis}%
\label{sec:sinfony_numres}

As the design approach for semantic communication and to solve~\eqref{eq:ce_opt} with the model~\eqref{eq:_sinfony_design_model}, we use our ML-based \sinfoni{} approach from~\cite{beck_semantic_2023,beck_model-free_2024}. However, we note that the conclusions derived from the results about the interplay between semantic communication and HDM models are not limited to this approach. These also extend to other methods,~e.g., model-based approaches, capable of providing the same quality of soft information at inference runtime. For example, \sinfonirl{} leverages reinforcement learning to train the design via~\eqref{eq:ce_opt} to comparable performance~\cite{beck_model-free_2024}.

\subsubsection{\sinfoni{} Design}

As shown in~\cite{beck_semantic_2023,beck_model-free_2024}, we apply \sinfoni{} to a distributed multipoint scenario, where meaning from multiple image sources is communicated to a single receiver for semantic recovery of the \rv{} $\relvec$. In the numerical example of~\cite{beck_semantic_2023}, four distributed agents extract features from different image views with an encoder based on the famous and powerful ResNet architecture~\cite{he_identity_2016} for rate-efficient transmission. Based on the received signals, the decoder recovers semantics by classification. Numerical results of~\cite{beck_semantic_2023} on images from the MNIST and CIFAR10 datasets show that \sinfoni{} outperforms classical digital communication systems in terms of bandwidth, latency and power efficiency.

In this article, we reuse the \sinfoni{} approach for integration with the HDM model. \sinfoni{} is particularly well-suited for integration because it can be easily adapted to any semantic source $\prob(\semvec,\relvec)$,~i.e., use case, including tool damage classification, by changing the data samples and specifically designing its DNN architecture.

In the guiding tool example, two image sensors provide different views of the tool (see \reffig{fig0:e2e_tool_example}). This results in a \sinfoni{} design (see Communications Design in \reffig{fig1:joint_sysmodel_human}) with two encoders $\prob_{\txpars}^{\indi}(\xvec_{\indi}|\semvec_{\indi})$ with $\indi=\{1,2\}$ that can be concatenated into one virtual encoder $\prob_{\txpars}(\xvec|\semvec)$, and one decoder $\aprob_{\rxpars}(\relvec|\yvec)$. When the image dimensions of $\semvec_{\indi}$ grow large, as in the case of Tool Wear, we improve dimension reduction of the \sinfoni{} encoder for low-complex and bandwidth-efficient feature extraction, adopting the ImageNet version of ResNet18~\cite{he_deep_2016},~e.g., including an initial extra Max Pooling step. This results in $\Nfeat=512$ features per encoder.

We test two \sinfoni{} Tx module designs that map those features onto the transmit signal $\xvec_{\indi}\in\xset^{\ntx\times 1}$: one with feature compression ($\ntx=128$) and one without ($\ntx=512$). Note that the number of channel uses $\ntx$ is proportional to bandwidth,~i.e., $\bandwidth\sim\ntx$. The signals $\xvec_{\indi}$ are transmitted over a communication channel $\prob(\yvec_{\indi}|\xvec_{\indi})$ to the decoder that consists of a common Rx layer of width $\nrx=1024$ processing the concatenated received signals $\yvec_{\indi}\in\yset^{\Nrx\times 1}$, each of length $\Nrx=\ntx$, and a final softmax layer with $\Nclass=2$ classes.

As in~\cite{beck_semantic_2023}, we train for $\snrtrain\in[-4,6]$~dB in model~\eqref{eq:_sinfony_design_model}. We also reuse the \sinfoni{} designs for the MNIST and CIFAR10 datasets from~\cite{beck_semantic_2023} and combine them with the HDM model. For the audio datasets, we use the same design from~\cite{beck_semantic_2023} but adopt the ImageNet version of ResNet18 because of the large spectrogram dimensions provided in \reftable{tab:datasets} as well. The version with feature compression is $\ntx=64$ and without $\ntx=512$. The source code including all details is available in~\cite{Beck_Semantic_INFOrmation_traNsmission_2024}.

\subsubsection{Communication Channel Model}

The proposed \sinfoni{} approach and likewise the proposed end-to-end sensing-decision framework are very generic in its assumptions about the communication channel $\prob(\yvec_{\indi}|\xvec_{\indi})$. For simplistic abstraction of reality and to study the interplay between semantic communication and human decision-making, the focus of this work, we assume an AWGN channel in the numerical experiments following~\cite{beck_semantic_2023}. In practice, many more channel effects such as multi-path interference, fading, and hardware non-idealities such as carrier frequency offset and phase offset have to be considered. In the two exemplary works~\cite{dorner_deep_2018,felix_ofdm-autoencoder_2018}, the authors show how a complete AE design of classical digital bit communication handling these effects can be achieved by means of ML-techniques, competitive or superior to expert-designed baseline approaches. These methods can be readily transferred to the \sinfoni{} design with its semantic-oriented transmit signals to account for more complicated channel effects, as they compensate the effects upfront bit reconstruction.

\subsubsection{\sinfoni{}-based Decision-Making}%
\label{sec:sinfony_alone}

\begin{figure*}[!t]%
    \centering
    \input{./TikZ/plot_sinfony_classic_new_humanrover_unnormalized.tikz}
    \caption{Comparison of the classification error rate of \sinfoni{} with different number of channel uses $\ntx$ per encoder and central image processing with digital image transmission on the Tool Wear, Speech Commands, and UrbanSound8K -- Fold 10 validation dataset as a function of SNR.}%
    \label{fig:sem_tools_classic_humanrover}
\end{figure*}

First, we evaluate the performance of \sinfoni{}-based decision-making within the semantic communication operation mode model~\eqref{eq:sinfony_e2e_model}, where \sinfoni{} makes the final decision via~\eqref{eq:sinfony_decision}.

\textbf{Tool Wear dataset:} \reffig{fig:sem_tools_classic_humanrover} shows the classification error rate $1-\acc$, with $\acc$ from~\eqref{eq:accuracy}, on the Tool Wear dataset over an AWGN channel as a function of SNR\@. The key findings are similar to those for MNIST and CIFAR10~\cite{beck_semantic_2023} but become much more obvious: Using less channel uses per encoder with \sinfoni{} $\ntx=128$ than the number of features ($\Nfeat=512$) results in the same performance compared to \sinfoni{} \txrx{} $\ntx=512$. This indicates that feature compression and thus a reduction in bandwidth is possible.

Moreover, we compare to central image processing by a ResNet classifier~\cite{Beck_Semantic_INFOrmation_traNsmission_2024} with classic digital transmission of the sensed images (\centralimagecommhuman{}): We assume that the RGB image bits are Huffman encoded, protected by an LDPC code with rate $0.25$ and BPSK modulated. At the receiver side, we use belief propagation for decoding. On average, the channel is utilized over $23,400$ times more frequently per encoder, with $\ntx\approx2,998,626.82\approx 3\cdot 10^6$ uses. Furthermore, at low SNR, significantly more power is needed to achieve the same classification error rate,~e.g., about $10$ dB more for $35\%$. Instead of graceful degradation as for \sinfoni{}, we observe a cliff effect typical for digital communication at a SNR threshold of $-2.5$ dB: Communication quality remains robust as long as channel capacity exceeds code rate and the LDPC code operates within its working point, but rapidly breaks down otherwise. This sharp contrast in performance and bandwidth highlights the huge potential of semantic communication.

\textbf{Audio datasets:} Applying \sinfoni{} for the first time to audio stimuli,~i.e., the Speech Commands and UrbanSound8K datasets, we observe the same qualitative behavior like for MNIST and CIFAR10. We have a SNR gap of $3$ dB, if we compress the features for transmission from $\ntx=512$ to $\ntx=64$. Comparing to $\ntx \approx 1.76 \cdot 10^5$ and $\ntx \approx 8.2 \cdot 10^5$ channel uses per audio sensor for Speech Commands and UrbanSound8K with digital communication of the 16-bit PCM signals, respectively, the reduction in bandwidth is substantial, consistent with the savings observed for Tool Wear. Moreover, the maximum accuracy of digital audio transmission slightly improves over distributed \sinfoni{} processing. This is most likely because of the artificial split of the audio data for distributed \sinfoni{} processing, such that correlations and patterns among splits are partly lost after processing. The real multi-view Tool Wear dataset does not exhibit this behavior, since it is not affected by any artificial splitting; thus, the observation is irrelevant for real-world scenarios.

\subsection{End-to-end Sensing-Decision Analysis}%
\label{sec:end2end_simulation}

Now, we assume our end-to-end sensing-decision model,~i.e., the overall model~\eqref{eq:e2e_model}: The semantic information in the images is transmitted by \sinfoni{} over a communication channel and then fed into the HDM model,~i.e., the GCM\@. Note that, in contrast to the \sinfoni{} scenario of \refsec{sec:sinfony_alone}, the HDM model now makes the final decision.

\subsubsection{Semantics Presentation Design}%
\label{sec:presentation_design}

In \refsec{sec:presentation_optimization}, we derive two design criteria for the semantics presentation $\hdecvec$: 1) It should keep all relevant information about $\relvec$. 2) It should fit to cognitive processing capabilities,~i.e., capacity, of the HDM model. We note that since HDM models are not capable to process the raw data of the images $\semvec$ directly~\cite{serre2016models} as outlined in \refsec{sec:presentation_optimization}, we cannot simply compare to a setup where the raw data of the images $\semvec$ are digitally communicated and processed. This means the design choice $\hdecvec=\semvec$ is ruled out in this work.

Therefore, we aim to gain first insights on the design trade-off by comparing HDM model performance with different presentations that reflect a different weighting of the two design criteria. To reflect practical considerations as outlined in \refsec{sec:presentation_optimization}, we design the presentation handcrafted based on the semantic decoder (see \reffig{fig0:e2e_tool_example}). In this context, in other words, we investigate how to balance the feature extraction of semantic communication and HDM models to achieve the best task performance,~i.e., to minimize the classification error rate.

We present either the categorical probability outputs (\prescat{}) or the relevant decision features (\presfeat{}) from \sinfoni{} as $\hdecvec$ to the HDM model,~i.e., the GCM:
\begin{enumerate}
    \item \textbf{\prescat{}:} The low-dimensional and interpretable probability estimate of \sinfoni{} for each category ($\hdecvec$-categorical) that fulfills design rule 2),~e.g., whether the tool is damaged or not:
          \begin{align}
              \hdecvec=\func[1]{\aprob_{\rxpars}(\relvec|\yvec)}=\begin{bmatrix}\aprob_{\rxpars}\left(\relvec={\ohfunc(1)}^{\trapo}|\yvec\right)\\ \vdots \\ \aprob_{\rxpars}\left(\relvec={\ohfunc(\Nclass)}^{\trapo}|\yvec\right)\end{bmatrix} \eqpoint
          \end{align}
    \item \textbf{\presfeatbog{}:} To provide the HDM model at an abstract level with potentially more relevant semantic information about $\relvec$ according to data processing inequality~\eqref{eq:data_processing_inequality} and design rule 1) for decision-making, we use the relevant decision features
          \begin{align}
              \hdecvec=\func[2]{\yvec}=\hidlay_{\aprob_{\rxpars}}^{(\NL-1)}(\yvec) \eqcomma
              \label{eq:relevant_features}
          \end{align}
          where $\hidlay_{\aprob_{\rxpars}}^{(\indhidlay)}\in\realnum^{\Nh^{(\indhidlay)}}$ is the output of the $\indhidlay$-th layer of $\aprob_{\rxpars}(\relvec|\yvec)$ with hidden layer width $\Nh^{(\indhidlay)}$ and the depth $\NL$ of the DNN\@. This means we extract the inputs to the final dense softmax layer of the \sinfoni{} decoder used for probability estimation, similar to a previous study that aims to model categorization with natural material~\cite{battleday_capturing_2020}.

          To further vary the level of detail or dimension of the presentation, we extract the most important $\Nfinalfeat=\{5,10,20,40,\Nh\}$ of the $\Nh=\Nh^{(\NL-1)}$ final layer features to facilitate effective processing of the HDM model according to design rule 2). The selection operation is defined by:
          \begin{subequations}
              \begin{align}
                  \hdecvec                         & =\modelfeaturesvec=\left[\hidlay_{\aprob_{\rxpars}}^{(\NL-1)}\right]_{\indi\in\Nselectedset}                                         \\
                  \Nselectedset                    & = \left\{\permutationel_{1},\permutationel_{2},\dots,\permutationel_{\Nfinalfeat}\right\} \quad \txt{with} \quad   \Nfinalfeat\le\Nh \\
                  \importance_{\permutationel_{1}} & \ge\importance_{\permutationel_{2}}\ge\dots\ge\importance_{\permutationel_{\Nh}}                                                     \\
                  \importancevec                   & =\abs[\wnnmat^{(\NL)}]^{\trapo}\cdot \ones_{\Nclass}\in \realnum^{\Nh\times 1} \eqpoint
              \end{align}
              \label{eq:selection_features}
          \end{subequations}
          The importance of the $\indi$-th feature ${[\hidlay_{\aprob_{\rxpars}}^{(\NL-1)}]}_{\indi}$, with respect to all output nodes $\aprob_{\rxpars}(\relvec={\ohfunc(\indc)}^{\trapo}|\yvec)$, is quantified by the importance vector $\importancevec$,~i.e., the sum of the absolute weight values in each column of the last-layer weight matrix $\wnnmat^{(\NL)}$.
\end{enumerate}
Based on the selected presentation,~i.e., \sinfoni{} features, the GCM classifies into $\Nclass$ categories,~e.g., into the binary tool states of wear and non-wear.

To present the \sinfoni{} features in the numerical evaluation, we use the \sinfoni{} version with $\ntx=128$ from \reffig{fig:sem_tools_classic_humanrover} for semantic communication (see \refsec{sec:sinfony_numres}) on the Tool Wear dataset, and the version with $\ntx=64$ on the Speech Commands and UrbanSound8K dataset. For MNIST and CIFAR10 datasets, we use the trained \sinfoni{} versions with $\ntx=56$ and $\ntx=64$ from~\cite{beck_semantic_2023, Beck_Semantic_INFOrmation_traNsmission_2024}. Note that $\ntx$ differs per dataset, since we tailored the \sinfoni{} architecture to the specific dataset. This holds also for the total number of final layer features which is $\Nh=\{56,64,1024,512,512\}$ for datasets MNIST, CIFAR10, Tool Wear, Speech Commands, and UrbanSound8K, respectively.

\subsubsection{Simulation Scenarios}
\label{sec:simulation_scenarios}

We evaluate our proposed framework on five datasets: Tools, MNIST, CIFAR10, Speech Commands and UrbanSound8K.

\textbf{Training Details:} The GCM parameters $\paramsgcm$ from~\eqref{eq:gcm_similarity} are optimized to maximize the empirical amortized cross-entropy~\eqref{eq:ce_gcm_opt} on the training set. We choose \sinfoni{}'s training SNR which is uniformly distributed between $[-4,6]$ dB. Furthermore, we set the learning rate to relatively high $\lr=1$ and train for one epoch $\Ne=1$ with batch size $\Nb=64$, as this leads to training convergence.

Moreover, we perform four main simulations --- a simulation of the accuracy as a function of the SNR typical for communications, including a joint optimization of semantic communication and HDM model, a simulation of the expertise, and a simulation of processing capacity of the HDM model both touching a psychological aspect:
\begin{enumerate}
    \item \textbf{SNR:} In the SNR simulation, we assume that the HDM model~\eqref{eq:gcm_model} has sufficient experience with the presented \sinfoni{} features,~i.e., it has perfect memory of the training set with $\datasetexemplar=\trainingset$,~i.e., the seen presented realizations encompass the entire training dataset of semantic communication. For evaluation, the HDM model receives the output of \sinfoni{} under varying SNR\@.
    \item \textbf{Expertise:} In the expert simulation, we assume the highest SNR of $20$ dB during communication for evaluation and vary the number of seen presentation realizations,~i.e., image/audio samples randomly selected from the training set. We define the number of seen realizations $\abs[\datasetexemplar]$ of classified tools as the expertise of the HDM model and simulate the performance at this expertise independently. Accordingly, a HDM model with high expertise has a larger knowledge base $\datasetexemplar$ compared to a HDM model with low expertise. The GCM parameters were optimized for the specific HDM knowledge bases $\datasetexemplar$.
    \item \textbf{Restricted Processing Capacity:} In the processing capacity simulation, we model a different factor of human cognitive ability: We simulate when the decision-maker is under stress or time pressure, which leads to reduced processing capacity during HDM~\cite{paul_input_2010},~e.g., resembling the capacity of the working memory,~i.e., how many stimuli can be held in consideration at once. Moreover, individual differences in working memory capacity and expertise exist, which affect the amount of information that can be accessed or processed.

          First, we train the GCM for the entire training set. During evaluation at the highest communication SNR of $20$ dB, however, we assume that the GCM only processes a certain subset of the presented relevant features under stress. To model this effect, we set some attention weights in $\simweightvec$ to zero and the remaining attention weights of the processed features are normalized, leading to stress-transformed weights $\simweightstressvec=\func[]{\simweightvec}$. Hence, we measure processing capacity by the zero-norm $\processingcapacity\le\norm{\simweightvec}_{0}$.

          We simulate processing capacities over the full range $\norm{\simweightvec}_{0}\le\Nfinalfeat$, but restrict the analysis to $\norm{\simweightvec}_{0}\in\{0,\dots,10\}$, since humans can only retain $2-9$ items in their short-term memory at once~\cite{cowan_magical_2001}. Furthermore, we assume that the HDM model selects the processed presentation variables according to their relevance,~i.e., the size of the attention weights in $\simweightvec$. This can be mathematically formulated, replacing the importance vector in~\eqref{eq:selection_features} by $\simweightvec$ and applying the selection matrix to $\simweightvec$. We choose either the most relevant features or the least relevant presented features to cover extreme cases.
    \item \textbf{Alternating Semantic Communication and HDM Training:} We investigate direct end-to-end optimization of semantic communication for HDM. This idea is enabled by our extension~\eqref{eq:ce_joint_opt} of the information-theoretic framework on semantic communication via~\eqref{eq:ce_presentation} from \refsec{sec:infomax_extension}. For a first investigation and fair comparison, we choose the \sinfoni{} and handcrafted presentation designs from before as a starting point.

          Since training of a human requires building a stable knowledge base,~i.e., based on fixed presentations, concurrent optimization of all parameters in batches does not resemble a practical scenario. Furthermore, we observe similar results of joint concurrent optimization compared to those of alternating training. Hence, we realize joint optimization via~\eqref{eq:ce_joint_opt} alternating $20$ times between overall encoder and HDM model training of $1$ epoch each. We train \sinfoni{} given the fixed HDM model by SGD of step size $\lr=10^{-3}$ consistent with the original learning rate schedule from~\cite{beck_semantic_2023}.
\end{enumerate}
In all simulations, the accuracy was calculated based on the validation dataset. We performed multiple Monte Carlo runs for evaluation: For all simulations, we iterated ten times through the dataset for each value of SNR, expertise level, and processing capacity, respectively. For the Tool Wear dataset, we iterated $100$ times.

\subsubsection{Numerical Results}%
\label{sec:num_res}

We present the simulation results in \reffig{fig:results_sinfony_gcm}, comparing to \sinfoni{}-based decision-making (\sinfoni{}) of the technical system on $\hrelvectech$ via~\eqref{eq:sinfony_decision} as the baseline. First, we note that the accuracy of the GCM is slightly worse than that of \sinfoni{}. This can be explained by the probabilistic decision process~\eqref{eq:gcm_decision} of the HDM model, which deviates from the optimal strategy to choose the most likely option. As discussed in \refsec{sec:hdm_optimization}, we note that HDM training rarely reaches full convergence in real environments, such that the simulations rather provide an upper bound on HDM performance~\cite{meagher_testing_2023}.

\begin{figure*}
    \centering
    \input{TikZ/plot_sinfony_hdm_horizontal_sinfony_curve_v2.tikz}
    \caption{The simulated classification performance of the proposed end-to-end sensing-decision framework, including \sinfoni{} and human decisions modeled by the GCM\@. Each column shows the accuracy on different datasets. From left to right: Tools, MNIST, and CIFAR10. The top row shows the accuracy as a function of SNR\@. The bottom row shows the accuracy as a function of the knowledge base size. Within each figure, the color of the lines indicates the type and number of features presented to the GCM.}%
    \label{fig:results_sinfony_gcm}
\end{figure*}

\subsubsection*{SNR Simulation}

Furthermore, the SNR curves show a similar trend for all three datasets. Accuracy increases as a function of SNR and plateaus after a certain SNR is reached. The performance of the HDM model is best when receiving the categorical probability input of \sinfoni{} (\prescat{}) compared to receiving the $\Nfinalfeat$ most important final layer features (\presfeat{}).

With the latter input, more features first yield better accuracy up to $\Nfinalfeat=40$, and the performance saturates after a certain number of feature dimensions. The number of features needed to reach saturation varies depending on the dataset and \sinfoni{} model. After $\Nfinalfeat=40$ the behavior is ambiguous: \presfeatN{\Nh} performs slightly better than $\Nfinalfeat=40$ for MNIST and more apparently for CIFAR10. However, accuracy deteriorates for Tool Wear.

This means that GCM is not always capable to learn to effectively extract the semantic information when provided with extra information, such as with $\Nfinalfeat=\Nh$. We assume that this behavior can be explained by overfitting since training converged. It is consistent with the bias-variance trade-off from statistical learning, which explains why low-capacity models generalize better with limited data~\cite{simeone2018brief}: GCMs with fewer parameters constrain the hypothesis space of solutions, effectively regularizing the learning process. In contrast, high-capacity GCMs with more inputs and attention weights tend to overfit to a limited knowledge base $\datasetexemplar$, based on a few seen realizations,~e.g., Tool Wear training data. We conclude that providing more features,~i.e., details, to the decision-maker with a small knowledge base $\datasetexemplar$ can lead to a suboptimal decision compared to providing less information.

In summary, the observed saturation suggests that the HDM model cannot extract additional relevant information about $\relvec$ from many feature inputs,~i.e., from a more detailed representation. In contrast, the model performs better with \sinfoni{}'s preprocessed probability estimates, highlighting \sinfoni{}'s ability to efficiently extract the semantic information. This finding implies that InfoMax-optimized outputs are well suited as inputs for human decision-making: \sinfoni{}'s graceful degradation translates directly into the GCM curve, and its accuracy effectively defines an upper bound for end-to-end performance. Varying the number of channel uses $\ntx$ merely shifts this upper bound without affecting the qualitative behavior, and we therefore omit a separate bandwidth-impact analysis.

\subsubsection*{Expertise Simulation}

The expertise simulation (bottom row in \reffig{fig:results_sinfony_gcm}) shows that accuracy increases with the number of seen presentation realizations and eventually plateaus. At high expertise starting around $\abstxt[\datasetexemplar]=10^3$, using the probability output of \sinfoni{} again results in the best performance compared to receiving the $\Nfinalfeat$ important feature dimensions. This shows that the GCM's semantic information processing was not as effective as that of \sinfoni{}.

However, at low expertise levels on the order of $100$, we observe the opposite --- providing last layer features leads to improved accuracy. We assume that the last-layer features, residing in a high-dimensional space, preserve structural information from \sinfoni{}'s training on the full dataset, including cross-class similarity relationships that the GCM can exploit. By contrast, \sinfoni{}'s class probabilities collapse these relationships into discrete, low-dimensional decisions. However, with enough training samples, the GCM is able to exploit the accurate \sinfoni{} predictions.

Moreover, like in the SNR simulation for $\Nfinalfeat=\Nh$, the accuracy does not always increase with the number of features $\Nfinalfeat$ under different expertise levels. For example, the accuracy on the Tool Wear dataset with $\Nfinalfeat=5$ and $10$ features exceeds the accuracy simulated with $20$, $40$ features at lower expertise, while performance with $\Nh$ is worst. For CIFAR10, the accuracy with $40$ features beats that with $\Nh$ features.

The observed behavior resembles that observed and explained in the SNR simulation. Hence, we attribute this behavior to overfitting of the GCM. The argumentation is supported by even smaller knowledge bases $\datasetexemplar$ compared to the full Tool Wear knowledge base example, making overfitting more likely and rendering fewer provided features more beneficial.

\subsubsection*{Audio Stimuli}

\begin{figure}[!t]
    \centering
    \input{TikZ/plot_sinfony_hdm_audio.tikz}
    \caption{The simulated classification performance of the proposed end-to-end sensing-decision framework on the audio datasets Speech Commands and UrbanSound8K -- Fold 10. The top row shows the accuracy as a function of SNR\@. The bottom row shows the accuracy as a function of the knowledge base size. Within each figure, the color of the lines indicates the type and number of features presented to the GCM.}%
    \label{fig:results_sinfony_gcm_audio}
\end{figure}

The simulation results of application of our end-to-end framework to audio stimuli,~i.e., on the Speech Commands and UrbanSound8k dataset, are shown in \reffig{fig:results_sinfony_gcm_audio}. The main observations and key conclusions align with those on the image datasets for both SNR and expertise simulation: For example, accuracy increases with SNR until it plateaus, \prescat{} leads to maximum performance and accuracy deteriorates using all $\Nfinalfeat=512$ last-layer features compared to using $\Nfinalfeat=40$ like in the case of Tool Wear. We note that the maximum number of presented features for the considered datasets grows very large to $1024$ or $512$ compared to $56$ or $64$ for MNIST and CIFAFR10, respectively. This supports the conclusion that overfitting is likely the cause.

We conclude that our findings on the end-to-end framework are transferable to other modalities, such as audio signals.

\subsubsection*{Restricted Processing Capacity Simulation}

\begin{figure*}[p]
    \centering
    \input{TikZ/plot_sinfony_hdm_horizontal_working_memory.tikz}
    \caption{The restricted processing capacity simulation of the proposed end-to-end sensing-decision framework, including \sinfoni{} and human decisions modeled by the GCM\@. Each plot shows the classification accuracy on one of the considered datasets as a function of the processing capacity $\processingcapacity$. Positive values correspond to selecting the $\processingcapacity$-nth most important features, while negative values correspond to selecting the $\processingcapacity$-nth least important features. Within each figure, the color of the lines indicates the type and number of features presented to the GCM.}%
    \label{fig:results_working_memory}
\end{figure*}

\reffig{fig:results_working_memory} shows the simulation results when the processing capacity of the HDM is restricted to $\processingcapacity\in\{0,\dots,10\}$, reflecting the typical human short-term memory capacity of $2-9$ items~\cite{cowan_magical_2001}. Positive and negative processing capacity values correspond to the two considered extreme cases of selecting the $\processingcapacity$-th most or least important decision features, respectively. This restriction introduces a discrepancy between training and testing environments, which can arise from multitasking, time pressure, or stress~\cite{paul_input_2010}, and can affect HDM performance.

Three effects can be observed in \reffig{fig:results_working_memory}:
\begin{enumerate}
    \item Depending on the dataset, the discrepancy can have little effect: When trained with $\Nfinalfeat=\{5,10,40\}$ on MNIST and tested for $\processingcapacity=5$ most relevant features, the performance difference is negligible. However, accuracy reduces drastically,~e.g., on Speech Commands comparing $\Nfinalfeat=\{10,40\}$ for $\processingcapacity=5$.

          Assuming that GCM training results in the same presented features to be weighted high, as the most relevant are shared across different feature numbers $\Nfinalfeat$, we expect that the performance degradation originates from the suboptimal balancing of the attention weights during testing. Thus, the distortion of the attention weights is higher when the discrepancy between training and testing is larger, and vice versa.
    \item Whenever GCM accuracy based on decision features (\presfeat{}) improves upon the V-shaped curve of processing capacity-restricted \sinfoni{} probability outputs (\prescat{}) where some probability values are simply not considered, we benefit from raw feature processing. This is the case on MNIST, Speech Commands, and UrbanSound8K.
    \item With restricted processing capacity selecting the least important features, whose number is depicted by negative values in \reffig{fig:results_working_memory}, accuracy improves with less presented features and maximizes for \presfeatN{5}. This is because with a focused pre-selection of important features, there are fewer choices to make mistakes, rendering this design choice most resilient to HDM mistakes.
\end{enumerate}

In summary, the results highlight that training humans with more features might not always lead to better desired performance. While training with large number of features would lead to higher performance when the decision maker can process all the available information, it also creates a situation where a discrepancy would lead to a drastic reduction in performance.

\subsubsection*{Alternating Semantic Communication and HDM Training}

So far, we have examined how semantic communication affects the decisions of a HDM model through different presentation designs in theory and simulations. Now, we investigate optimizing semantic communication directly for the given HDM model via~\eqref{eq:ce_joint_opt} to improve decisions, as detailed in \refsec{sec:simulation_scenarios}.

The results in \reffig{fig:results_sinfony_gcm_joint_training} show that decision improvements are possible with End-to-End (E2E) training: For MNIST, even with \presfeatN{5}, we approach the maximum accuracy. However, for all other datasets, accuracy degrades with \presfeatN{5}, meaning that alternating training converges to a worse solution compared to separate subsequent training. With more features,~e.g., \presfeatN{20}, this trend reverses to consistent improvements except for Tool Wear. Notably, we can even improve beyond the best separate-training accuracy in the case of UrbanSound8K with \presfeatN{\numberlastlayerfeatures}. A large presentation dimension seems to be important for stable and effective alternating training.

Consistent with this observation is that the accuracy surprisingly deteriorates with \prescat{} --- for the smaller datasets Tool Wear and UrbanSound8K close to that of a random guess. This indicates that this presentation design is not well-suited for a joint optimization, especially with limited data. The softmax layer acts in \sinfoni{} joint optimization as an intermediate layer with known common drawbacks such as vanishing gradients due to saturation, projection into a lower-dimensional manifold of a simplex, numerical instability, and dense coupling of input variables.

In summary, the results are inconclusive: Significant accuracy gains are possible, but so are substantial losses. Given the substantial effort required to implement such a joint end-to-end optimization, its practical usefulness remains questionable. First, the HDM model has to be trained to \sinfoni{}, and vice versa, until training convergence. This adds multiple alternating training epochs and thus much training complexity. Second, if a HDM model is not used as a bio-inspired machine, repeated rounds of training of real humans to numerous \sinfoni{} training outputs,~i.e., $\Nsamp=60$ k for MNIST, can take quite long in practice. Third, computing gradients of real human decisions for \sinfoni{} optimization poses practical challenges, as elaborated in \refsec{sec:outlook}. More research is needed to clarify the potential of an end-to-end design.

\begin{figure*}[p]
    \centering
    \input{TikZ/plot_sinfony_hdm_joint_training_accuracy.tikz}
    \caption{The simulation results of End-to-End (E2E) training in the proposed end-to-end sensing-decision framework, including \sinfoni{} and human decisions modeled by the GCM\@. Each plot shows the classification accuracy on one of the considered datasets as a function of SNR. Within each figure, the color of the lines indicates the type and number of features presented to the GCM.}%
    \label{fig:results_sinfony_gcm_joint_training}
\end{figure*}

\subsubsection*{Discussion and Main Conclusions}

Recalling the design trade-off in~\eqref{eq:ce_presentation} from \refsec{sec:presentation_optimization}, we conclude from all simulation scenario observations that it is more important to match the cognitive capabilities of the GCM by a low-dimensional presentation, in the considered tasks than providing more relevant information about $\relvec$. Joint optimization can improve GCM accuracy. Moreover, while accuracy is maximum with more elaborate \sinfoni{} preprocessing,~i.e., the final \sinfoni{} output, few raw decision features prove to be useful under non-ideal conditions such as low expertise and low processing capacity.

A crucial benefit of a low-dimensional presentation is a reduction in HDM processing complexity, leading to faster decision-making. This means semantic communication can facilitate not only effective but also efficient HDM by matching to the cognitive process. Although cognitive models such as ACT-R can simulate the temporal dynamics of feature integration, they differ fundamentally from the HDM model considered here~\cite{ritter_actr_2019}. A detailed analysis of reaction times based on such frameworks is therefore beyond the scope of this work and left for future research.

Providing raw features can also offer several other benefits. First, there is a slight reduction in processing complexity, since some nodes are removed. Second, in this work, \sinfoni{} is optimized for a single task,~i.e., deciding whether a tool needs to be replaced. In more complex situations, however, HDM may need to deal with unexpected events or changing goals not covered by the current form of our end-to-end sensing-decision framework. A more detailed representation allows the HDM model to react flexibly to these changes, compared to relying solely on probability estimates. Furthermore, when human decision-makers are involved instead of simulated ones, using richer features could positively influence motivational and emotional factors, such as experienced autonomy and competence, potentially improving performance~\cite{ryan_self-regulation_2006}, as we will discuss in the outlook in \refsec{sec:outlook}.

\section{Answering the Research Questions}%

The main takeaways from the numerical simulations and theoretical investigations provide answers to the research questions stated in the beginning:
\begin{itemize}
    \item[a)] \emph{\textbf{Joint Modeling:} How to model the end-to-end sensing-decision-making process that bridges the disciplines communications and psychology?} Exploiting Markov chains and generative probabilistic models, we can model the end-to-end sensing-decision process by~\eqref{eq:e2e_model}. This framework is broadly applicable and can be naturally extended to incorporate sensing, timing, and power aspects.
    \item[b)] \emph{\textbf{Suitability \& Design:} Is semantic communication suitable for providing the information needed in terms of relevance and accuracy to facilitate effective HDM\@?} The semantic information provided by the \sinfoni{} architecture supports HDM, as motivated by the InfoMax principle. An end-to-end optimization has the potential to improve the HDM accuracy.

          \emph{Given a task, how to design semantic communication for accurate HDM, specifically:}
          \begin{itemize}
              \item[i)] \emph{Which information should be provided to HDM?}  Using raw decision features leads to imperfect information integration of the HDM model compared to \sinfoni{}, evident through saturation in the simulations. However, this depends on the context. For example, with little expertise or restricted processing capacity, providing few raw decision features can help the HDM model relying on similarity structure to extract more information compared to \sinfoni{}'s compressed probability outputs. This is consistent with the design trade-off derived from~\eqref{eq:theoretical_design_criteria} and shows that it is more important to match the cognitive capabilities of the GCM than simply providing more relevant information,~e.g., by presenting structural richer information when the knowledge base is small, and relying more on elaborate \sinfoni{} feature extraction with larger knowledge bases.
              \item[ii)] \emph{How much information should be provided to HDM?} Providing more detailed representations,~i.e., more features, does not always increase HDM decision accuracy and can decrease HDM processing efficiency. The saturation indicates that the HDM model at some point misses subtle details in the additional features and can be even misguided. This effect is more pronounced at little expertise and limited processing capacity. This context-dependence requires carefully balancing the design trade-off,~i.e., the information provided by semantic communication with the HDM process.
          \end{itemize}
    \item[c)] \emph{\textbf{Impact:} How does the HDM process impact the end-to-end sensing-decision-making process?} The accuracy of the HDM model can be inferior to that of semantic communication systems due to the probabilistic nature of HDM, non-ideal training, and possibly restricted processing capacity.
\end{itemize}

\section{Outlook -- Open Questions and Challenges}%
\label{sec:outlook}

The proposed end-to-end sensing-decision framework is a first step towards integrating semantic communication and the human receiver. We will now explore remaining open questions and challenges with respect to all our framework components from \reffig{fig1:joint_sysmodel_human}, and what they mean for semantic communication.

\subsection{Challenge: Practical Optimization of Semantic Communication for Human Decisions}

In this article, we have examined how semantic communication can be optimized directly for the given human or HDM model to improve decisions in theory and simulations. The idea is to include the human or HDM model in an end-to-end optimization process of all framework parameters through~\eqref{eq:ce_joint_opt} and is supported by our extension of the information-theoretic framework on semantic communication via~\eqref{eq:ce_presentation} from \refsec{sec:infomax_extension}.

However, since both the human and the HDM model are essentially a black box that is not known or differentiable in practice as assumed in this work, this seems difficult. Nevertheless, it is possible to evaluate the cross-entropy loss or another target metric for the human or HDM model decisions and feed it back to \sinfoni{} as a reward. Thus, one idea could be to use the stochastic policy gradient as in~\cite{beck_model-free_2024} for \sinfoni{} optimization to allow joint alternating optimization over the whole chain, including \sinfoni{} and the human/HDM model. This means end-to-end optimization is a realistic approach but requires substantial more training iterations and time compared to the separated \sinfoni{} and subsequent GCM optimization.

\subsection{Challenge: Limitations of Human Decision-Making Models}

To include the HDM model in an optimization process, it is essential to have an accurate representation of the HDM process. Here, we simulated the decision-making process by applying a computational model, the GCM, for illustrative purposes. While our simulation reflects many traits of HDM with human participants, not all assumptions will apply to realistic categorization decisions. For instance, given the lack of HDM data for the simulated tasks, we assumed perfect memory and training convergence on the training set for the GCM\@. These assumptions are difficult to achieve in a real-life situations, but more appropriate assumptions are likely to strongly depend on individuals and tasks.

Accordingly, just like in most of psychological research on HDM, future HDM models will need to be carefully selected and designed to accurately reflect human decision processes for the task of interest,~e.g., by accounting for limitations in human information retrieval~\cite{nosofsky_exemplar-based_2015,stewart_decision_2006} and contextual influences on the decision process such as limited cognitive resources due to multitasking, acute stress, or time pressure~\cite{lamberts1995categorization, seitz2023testing}. As outlined in \refsec{sec:hdm_model}, it is possible to extend the basic GCM~\eqref{eq:gcm_model} used in this work to model many of these HDM aspects. Nevertheless, experimental validation with human participants, typical for psychological experiments, is required to develop and validate appropriate HDM models for the decision problem at hand, assess the beneficial effects of semantic communication and their visualizations, and understand the trade-offs between performance facilitation and potential negative impacts on motivation.

\subsection{Challenge: Presentation of Semantic Information}

For tractability in the simulations and to reflect constraints on realistic cognitive processing, we provided the HDM model with two interpretable, handcrafted presentations containing varying amount of information --- probability outputs or the most important decision features of semantic communication. The key question is how this abstract representation $\hdecvec=\func{\yvec}$ or process $\prob(\hdecvec|\yvec)$, shown in \reffig{fig1:joint_sysmodel_human}, translates into real-world scenarios with human subjects. To help humans interpret the output $\yvec$ or $\aprob_{\rxpars}(\relvec|\yvec)$, it is essential to present it through a Human-Machine Interface (HMI) that connects human users with semantic communication. In practice, this could involve visualizing,~e.g., tool damage probabilities, through symbols and colors on a screen or in augmented/virtual reality, with varying levels of detail (see \reffig{fig0:e2e_tool_example}) --- from basic machine learning outputs (e.g., tool wear status) to more detailed insights such as algorithm certainty, textual explanations, and even image-based class activation mapping~\cite{Lee_19}.

The HMI is a critical component, as the presentation format can strongly influence the HDM process~\cite{HAWLEY2008448, strathie2017presentation}. It must hence present complex information in a way that enables informed decisions while maintaining essential context. Designing a successful HMI requires an understanding of the domain in which it operates, including the industry, the user base, and the types of decisions being made~\cite{Nikolova_22}. A context-aware HMI that adapts to the user's history, preferences, and current situation can further improve decision-making by providing personalized and relevant information~\cite{Salima_23}.

\subsection{Challenge: Variability in Human Decision Goals and Expertise}

For many tasks, human decision-makers will differ in their preferences, intentions, and levels of expertise. In addition, task goals may be multi-faceted and subject to change of time, requiring to adjust transmitted information to the individual and the current situation. For example, when communicating information about tools, a person interested in understanding how different machines affect tool usability will need different information than someone focused on identifying tools that need to be replaced. In addition, an expert is likely to prefer a detailed, rich presentation, while a novice may benefit more from clear, concise support.

Furthermore, individual differences can lead to different information even when the decision objective is the same. For example, people differ in how much risk they are willing to accept~\cite{wang_comparison_2009}. Given the same information about the probability that the tool will fail, a risk-seeking person may conclude that the risk is acceptable, while a more risk-averse person would choose to exchange it. Thus, semantic communication that attempts to optimize tool use while keeping the failure rate below a tolerable threshold may require adapting the information conveyed to the decision-maker, for example by changing how potential risks are presented~\cite{torra_handling_2015, liesio_incomplete_2023}. While the core model~\eqref{eq:gcm_model} of the GCM investigated in this work does not accommodate all individual differences, it can be extended to simulate variability.

\subsection{Challenge: Conflict of Interest between Sender and Receiver}

The interests or goals of a human sender may not be well aligned with those of the human receiver. A fundamental factor contributing to such a misalignment of interests could be that human receivers are risk-averse. For example, even if a tool remains functional, the receivers may classify it as defective in order to avoid potential errors, since they are reluctant to take responsibility for using a worn tool. The sender thus has an incentive to manipulate the message in order to influence the receiver's decisions. If the difference in interests is too large, the receiver could ignore any message the sender sends.

This means that successful semantic communication also depends on trust between sender and receiver. Economists, following~\cite{Crawford1982}, have long studied this sender-receiver problem using game theory. For a recent overview of this literature, see~\cite{Blume2020}. They found that the amount of information that can be transmitted depends on how large the difference in interests is. Considering how much the sender wants to manipulate the information to influence the receiver's action is important in semantic communication. Even if the technology allows for very accurate transmission of semantic meaning, the best transmission strategy would still depend on the characteristics of the sender and receiver.

\section{Conclusion}

In this paper, integrating an interdisciplinary perspective from communications and psychology, we proposed a probabilistic end-to-end sensing-decision framework that wirelessly links sensed data with Human Decision-Making (HDM) by semantic communication. We analyzed this integration exemplarily using \sinfoni{} and an effective HDM model based on generalized context models for various datasets. The theoretical and numerical results indicate that semantic communication can optimize task performance by balancing information detail with human cognitive processes, achieving accurate decisions while demanding less bandwidth, power, and latency compared to classical methods.

This work is intended to inspire further interdisciplinary research on higher semantic levels of communication. Open questions include how to practically optimize semantic communication for human decisions, how to extend the HDM model, how to design human-machine interfaces that convey meaning more effectively, and how to account for different intentions between sender and receiver as well as individual differences among receivers.

\appendices{}

\section{Non-Convexity of the Cross-Entropy in the Generalized Context Model Parameters}
\label{sec:non_convexity}

We show that the objective function~$\Lossf_{\paramsgcm}^{\mathtxt{CE}}$ in~\eqref{eq:ce_gcm_opt} is non-convex with respect to the GCM parameters~$\paramsgcm$.
\begin{proof}
    First, in the GCM model~\eqref{eq:gcm_model} and~\eqref{eq:gcm_similarity}, the term $\func{\simconst,\simweightvec}=\simconst\cdot\distfunc{\hdecvecreal_{1},\hdecvecreal_{2}|\simweightvec}$ is not jointly convex in the similarity gradient $\simconst$ and attention weights $\simweightvec$, as the Hessian matrix
    \begin{align}
        \frac{\partial^2 \func{\simconst,\simweightvec}}{\partial^2 \left[\simconst,\simweightvec\right]^{\trapo}} & =
        \begin{bmatrix}
            0                               & \nabla \distfunc{\simweightvec}^{\trapo}                                            \\
            \nabla \distfunc{\simweightvec} & \simconst\cdot \frac{\partial^2 \distfunc{\simweightvec}}{\partial^2 \simweightvec}
        \end{bmatrix}
    \end{align}
    is indefinite for non-constant $\distfunc{\simweightvec}\neq\constvec$. This can be shown through the quadratic form
    \begin{align}
          & \begin{bmatrix}\plvar_{1} & \plvec_{2}\end{bmatrix} \cdot\frac{\partial^2 \func{\simconst,\simweightvec}}{\partial^2 \left[\simconst,\simweightvec\right]^{\trapo}} \cdot \begin{bmatrix}\plvar_{1}\\\plvec_{2}\end{bmatrix} \nonumber \\
        = & \; 2\cdot \plvar_{1} \cdot \nabla \distfunc{\simweightvec}^{\trapo} \cdot \plvec_{2} + \simconst\cdot\plvec_{2}^{\trapo} \cdot \frac{\partial^2 \distfunc{\simweightvec}}{\partial^2 \simweightvec} \cdot \plvec_{2}                   \\
        < & \; 0 \qquad \txt{for} \quad \plvar_{1}\to-\infty
    \end{align}
    that becomes negative for $\plvar_{1}\to-\infty$, for fixed $\plvec_{2}$ such that $\nabla \distfunc{\simweightvec}^{\trapo} \cdot \plvec_{2}\neq \zero$ and $\nabla \distfunc{\simweightvec}\neq 0$, considering that the second term is constant in $\plvar_{1}$. Since the objective function~$\Lossf_{\paramsgcm}^{\mathtxt{CE}}$ from~\eqref{eq:ce_gcm_opt} is composed of this non-convex function, it is non-convex in~$\paramsgcm=\{\simconst,\simweightvec\}$ itself as well.
\end{proof}

\bibliographystyle{IEEEtran}
\bibliography{IEEEabrv, ANT_references/bibliography.bib, ANT_references/bibliography_journalCMD.bib, ANT_references/bibliography_semantic.bib, ANT_references/bibliography_semantic_rl.bib, ANT_references/bibliography_introduction.bib, ANT_references/bibliography_ownpub.bib, ANT_references/bibliography_ownpub_other, ANT_references/bibliography_scil.bib, ANT_references/bibliography_humanrover.bib, references.bib}

\begin{IEEEbiography}[{\includegraphics[width=1in, height=1.25in, clip, keepaspectratio]{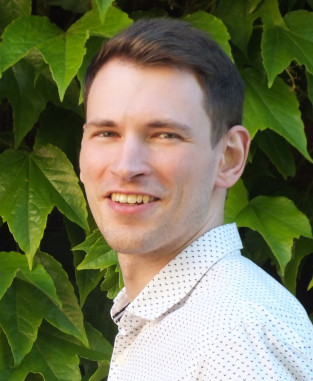}}]{Edgar Beck } (Graduate Student Member, IEEE) received the B.Sc.~and M.Sc.~degrees in electrical engineering from the University of Bremen, Germany, in 2014 and 2017, respectively, where he is currently pursuing a Ph.D. degree in electrical engineering at the Department of Communications Engineering (ANT). His research interests include cognitive radio, compressive sensing, massive MIMO systems, semantic communication, and machine learning for wireless communications.
    He was a recipient of the OHB Award for the best M.Sc.~degree in Electrical Engineering and Information Technology at the University of Bremen in 2017.
\end{IEEEbiography}

\begin{IEEEbiography}[{\includegraphics[width=1in, height=1.25in, clip, keepaspectratio]{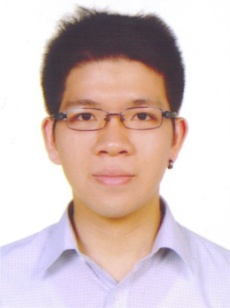}}]{Hsuan-Yu Lin } is a postdoctoral researcher at the Department of Psychology at the University of Bremen. He received his M.Sc.~degree in psychology from National Chengchi University in Taiwan and his Ph.D. in psychology from University of Zürich in Switzerland. His research focus is on decision-making processes, particularly patch-leaving and sampling decisions.
\end{IEEEbiography}

\begin{IEEEbiography}[{\includegraphics[width=1in, height=1.25in, clip, keepaspectratio]{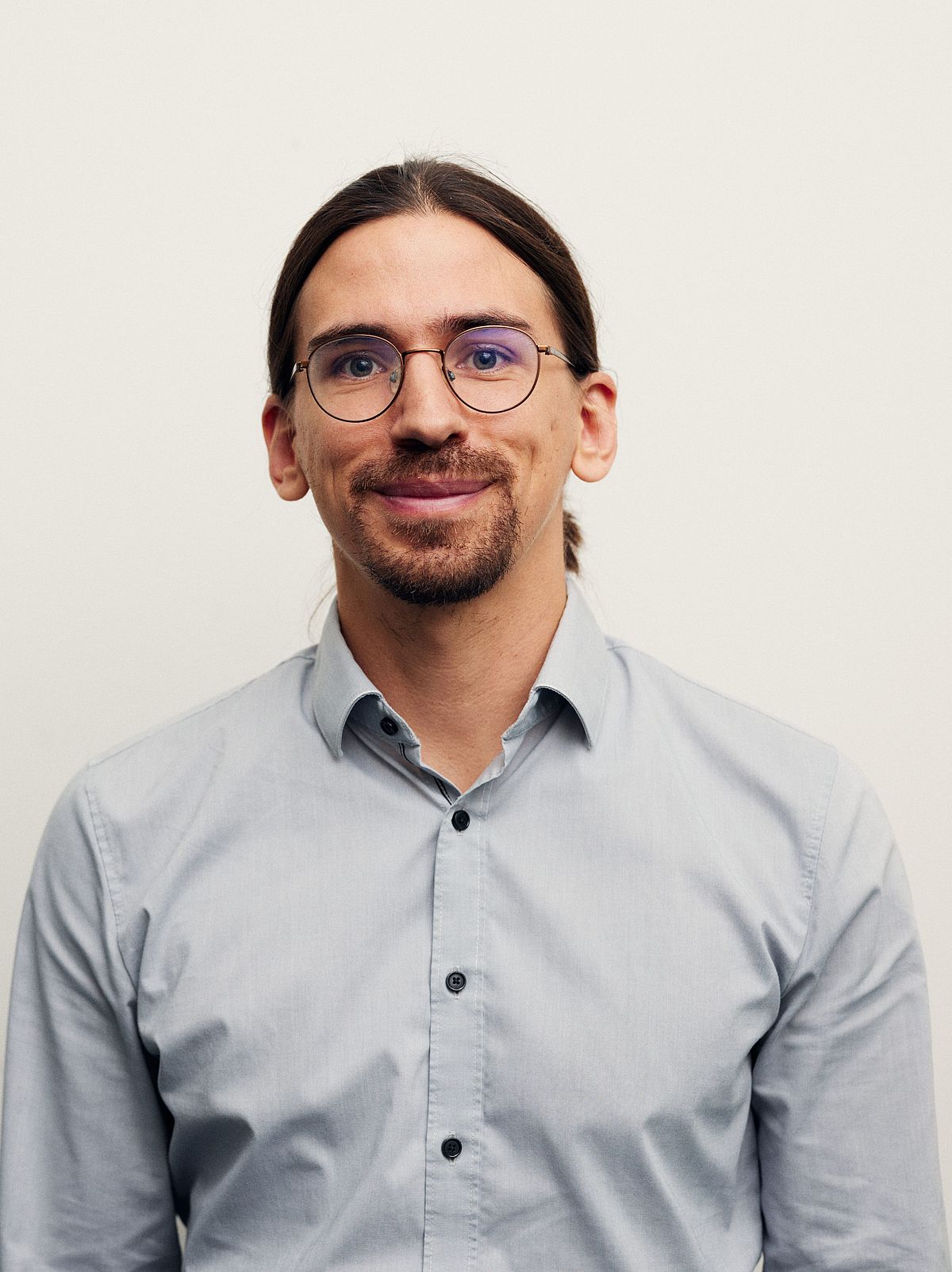}}]{Patrick Rückert } is chief engineer at the Bremen Institute for Mechanical Engineering at the University of Bremen. He received his B.Sc.~in industrial engineering from the University of Applied Sciences in Karlsruhe and his M.Sc.~in production engineering at the University of Bremen. His research focuses on human-robot collaboration, augmented and virtual reality systems, and the integration of machine learning for assembly processes.
\end{IEEEbiography}

\begin{IEEEbiography}[{\includegraphics[width=1in, height=1.25in, clip, keepaspectratio]{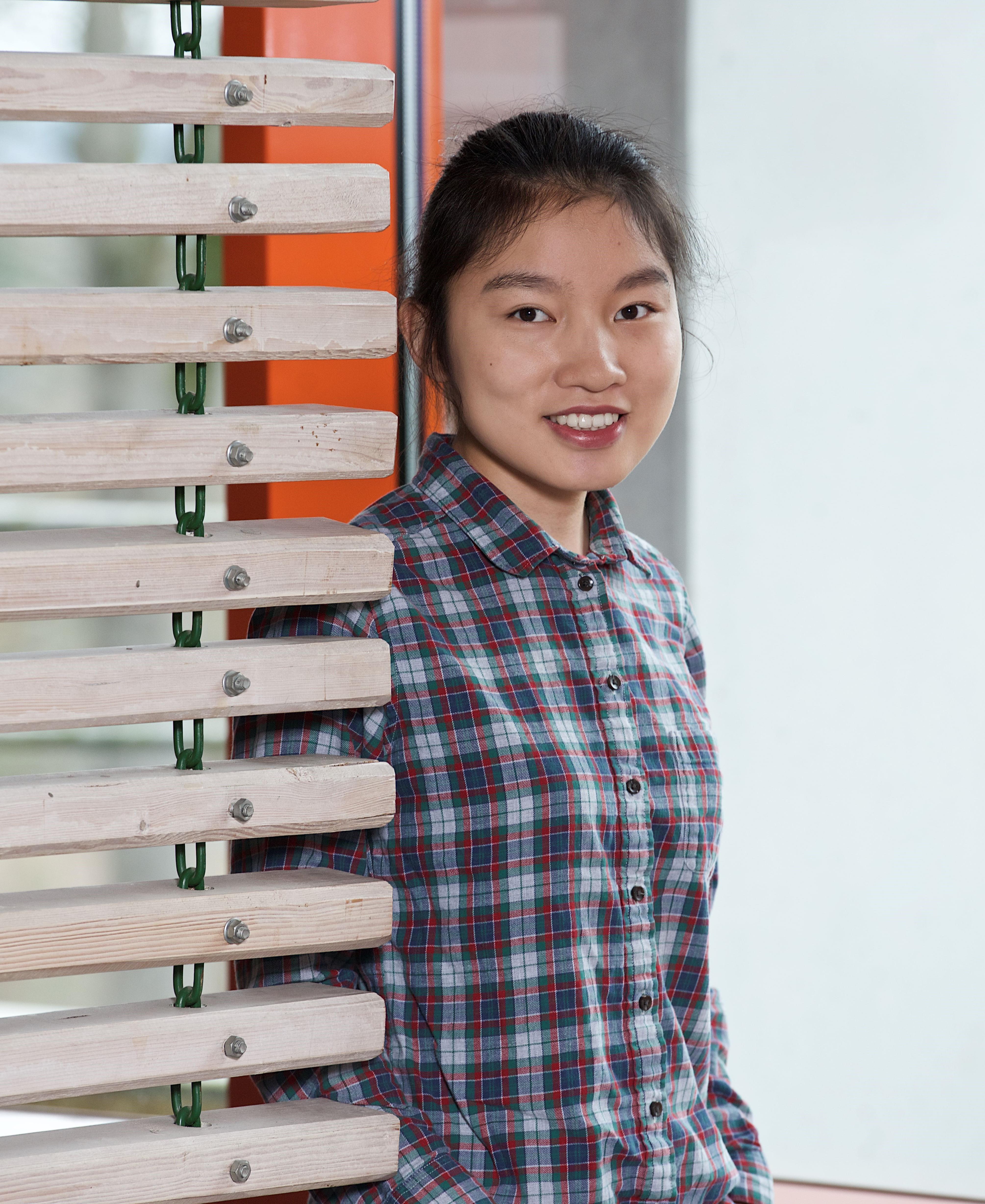}}]{Yongping Bao } is a postdoctoral researcher at the SOCIUM institute of the University of Bremen. She received her Ph.D. in Economics from the University of Konstanz. Her research focuses on behavioral and experimental economics, game theory, and human-algorithm interaction.
\end{IEEEbiography}

\begin{IEEEbiography}[{\includegraphics[width=1in, height=1.25in, clip, viewport=40 0 330 360, keepaspectratio]{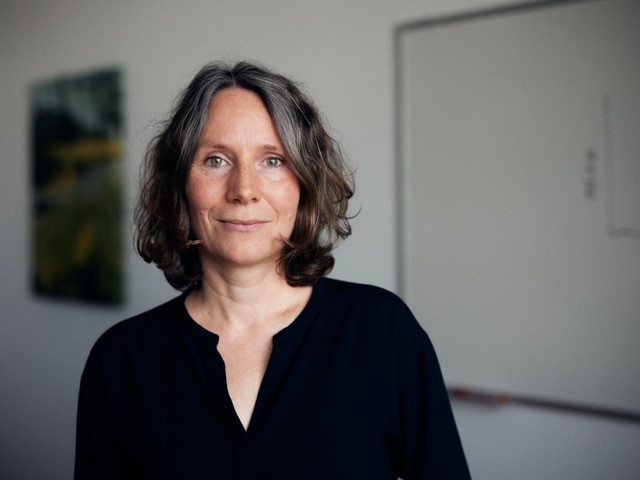}}]{Bettina von Helversen } is a professor for Experimental Psychology at the University of Bremen since 2019. Before, she held a Swiss Science Foundation Professorship at the University of Zurich for Cognitive Decision Psychology and worked at the Center for Economic Psychology in Basel. She received her Ph.D. from the Humboldt University in Berlin. She is interested in understanding and modeling how people make judgments and decisions. In her work she focuses on the different cognitive strategies people use to solve these tasks and the factors that influence strategy selection such as task structure, memory, affect or stress and how these change over the life span.
\end{IEEEbiography}

\begin{IEEEbiography}[{\includegraphics[width=1in, height=1.25in, clip, viewport=70 10 230 220, keepaspectratio]{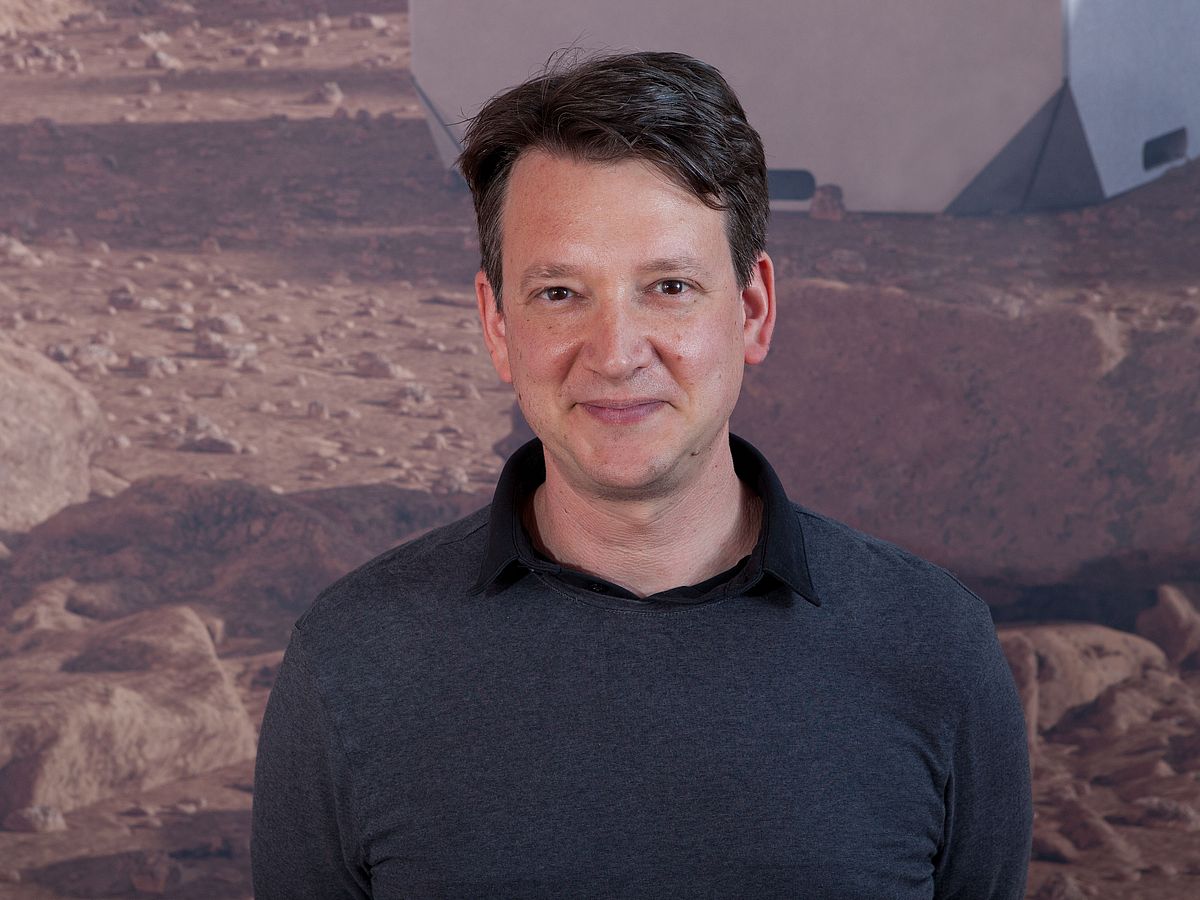}}]{Sebastian Fehrler } joined the University of Bremen, SOCIUM, as Professor of Economics in 2020. He studied Statistics (B.Sc., HU Berlin) and Economics (M.Sc., U Nottingham;  Ph.D., U Zurich), spent a year as a visiting postdoc at the Center for Experimental Social Sciences at NYU, and worked as an assistant professor at the University of Konstanz. His research interests are in the fields of public, organizational and behavioral economics. He uses game theory and experiments to address his research questions.
\end{IEEEbiography}

\begin{IEEEbiography}[{\includegraphics[width=1in, height=1.25in, clip, viewport=350 0 1050 900, keepaspectratio]{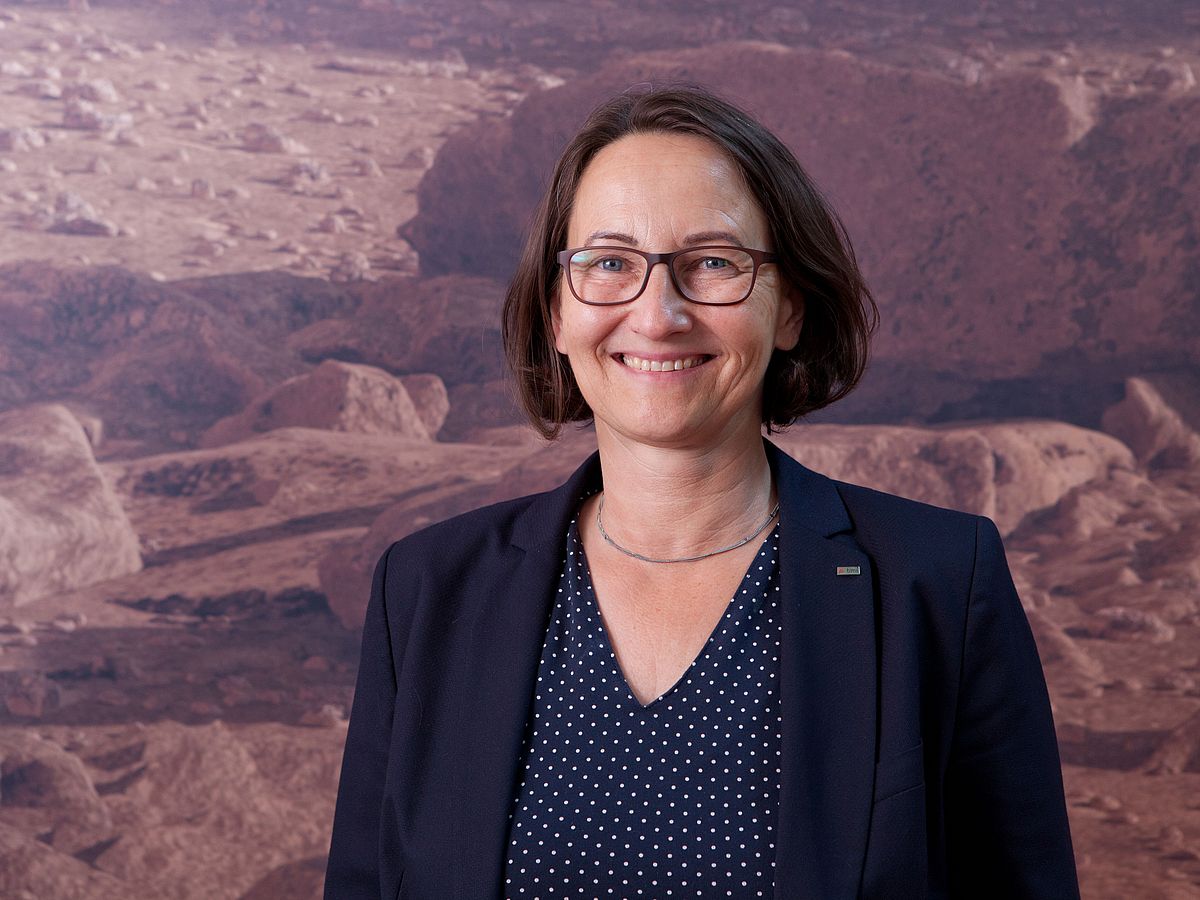}}]{Kirsten Tracht } is professor at the faculty of production engineering of the University of Bremen and speaker of the Bremen Institute for Mechanical Engineering. She studied mechanical engineering at the University of Hannover and received her Ph.D. under the supervision of Prof.~H.K.~Tönshoff in the field of die and mold making. She received the Kienzle-Medal. She is responsible for the master's study course `production engineering'. Her research focuses on assembly technologies and production design techniques in industrial settings.
\end{IEEEbiography}

\begin{IEEEbiography}[{\includegraphics[width=1in, height=1.25in, clip, keepaspectratio]{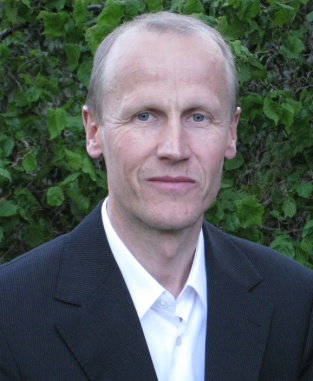}}]{Armin Dekorsy } (Senior Member, IEEE) is a professor at the University of Bremen, where he is the director of the Gauss-Olbers Space Technology Transfer Center and heads the Department of Communications Engineering. With over eleven years of industry experience, including distinguished research positions such as DMTS at Bell Labs and Research Coordinator Europe at Qualcomm, he has actively participated in more than 65 international research projects, with leadership roles in 17 of them. He is a Senior Member of the IEEE Communications and Signal Processing Society and a member of the VDE/ITG Expert Committee on Information and System Theory. He co-authored the textbook `Nachrichtenübertragung, Release 6, Springer Vieweg', which is a bestseller in the field of communication technologies in German-speaking countries. His research focuses on signal processing and wireless communications for 5G/6G, industrial radio, and 3D networks.
\end{IEEEbiography}

\end{document}